\documentclass[superscriptaddress, prd, aps,amsmath,amssymb,showpacs,showkeys, twocolumn]{revtex4-2}
\usepackage{graphicx}
\usepackage{xcolor}
\usepackage[%
  colorlinks=true,
  urlcolor=blue,
  linkcolor=red,
  citecolor=blue
]{hyperref}
\usepackage{orcidlink}

\begin{document}

\title{Quantum Oppenheimer-Snyder Black Holes with a Cloud of Strings Surrounded by\\[2mm] Perfect Fluid Dark Matter }

\author{Faizuddin Ahmed\orcidlink{0000-0003-2196-9622}}
\email{faizuddinahmed15@gmail.com}
\affiliation{Department of Physics, The Assam Royal Global University, Guwahati 781035, Assam, India}

\author{Allan R. P. Moreira\orcidlink{0000-0002-6535-493X}}
\email{allan.moreira@fisica.ufc.br (corresp. author)}
\affiliation{Secretaria da Educação do Ceará (SEDUC), Coordenadoria Regional de Desenvolvimento da Educação (CREDE 9), Horizonte, Ceará 62880-384, Brazil}

\author{Abdelmalek Bouzenada\orcidlink{0000-0002-3363-980X}}
\email{abdelmalekbouzenada@gmail.com}
\affiliation{Laboratory of Theoretical and Applied Physics, Echahid Cheikh Larbi Tebessi University, Algeria}
\affiliation{Research Center of Astrophysics and Cosmology, Khazar University, Azerbaijan}

\date{\today}

\begin{abstract}
In this study, we examine quantum Oppenheimer-Snyder black holes (BHs) embedded within a cloud of strings and immersed in perfect fluid dark matter. Also, beginning with the underlying spacetime geometry, we determine how quantum corrections, string cloud contributions, and dark matter effects alter the geometrical structure and physical characteristics of the BH. Also, the optical behavior is investigated via a systematic analysis of the photon sphere and the associated BH shadow, emphasizing possible observational features capable of differentiating this configuration from classical models. We also analyze the motion of test particles, focusing on how surrounding matter components affect trajectories, stability conditions, and effective potentials. Scalar field perturbations are considered to investigate the BH response to external excitations and to extract information regarding its dynamical properties. In this case, the thermodynamic behavior of the system is studied, including the role of string clouds and dark matter in modifying BH thermodynamic quantities. Also, the obtained results present a unified description of the combined effects of quantum corrections, nonstandard matter sources, and BH physics, with potential relevance for both observational constraints and theoretical modeling of compact objects.
\end{abstract}

\maketitle

\tableofcontents

\section{Introduction}\label{sec:1}

Black holes (BHs) constitute central predictions of Einstein’s general relativity (GR) \cite{GR2}, first defined in 1916 \cite{GR1}. Also, they arise from complete gravitational collapse, generating a space-time region from which no signal can escape, delimited by the event horizon (EH). A major advance occurred in 2015 when the LIGO Collaboration observed gravitational waves (GW) emitted during the merger of two stellar-mass black holes (BHs), providing direct validation of GR in the strong-field regime and establishing GW astronomy \cite{BH1}. In 2019, the Event Horizon Telescope (EHT) delivered the first horizon-scale image of a supermassive BH in $M87^{*}$, showing a bright asymmetric ring compatible with GR predictions for photon motion in the vicinity of an EH \cite{EHTL1, EHTL4, EHTL6}. More recently, EHT resolved Sagittarius $Sgr A^{*}$ at the center of the Milky Way, where the measured ring diameter was consistent with GR expectations despite variability \cite{EHTL12, EHTL14, EHTL15, EHTL16, EHTL17}. In this context, these results have established BHs as observational laboratories for probing strong-field gravity and high-energy astrophysical processes.

BH thermodynamics was introduced in \cite{CSR0}, where the study illustrates a generalized off-shell free energy formalism to categorize BHs into three distinct topological classes determined by their topological numbers. Also, this formalism enables a systematic investigation of universal thermodynamic characteristics across diverse BH solutions. In the present study, we extend this framework to stationary and axisymmetric BHs embedded in a perfect fluid dark matter (PFDM) environment. PFDM, characterized by an equation of state $p/\rho = \epsilon$, produces static spherically symmetric space-times that can be represented through power-law or logarithmic metric functions \cite{CSR1, CSR2, CSR3}. Its astrophysical significance has been explored through analyses of phantom fields constrained by stellar dynamics in spiral galaxies \cite{CSR4, CSR5}. The presence of PFDM modifies several BH properties, including shadows, quasinormal mode spectra, and gravitational lensing patterns \cite{CSR6, CSR7, CSR8, CSR9, CSR10, CSR11}. Extensions of topological thermodynamic analysis have been performed for charged ($Q$) and rotating BH solutions \cite{CSR13, CSR14, CSR15, CSR16, CSR17, CSR18, CSR19, CSR20, CSR21, CSR22, CSR23, CSR24, CSR25, CSR26, CSR27, CSR28}. In particular, logarithmic PFDM models have been extensively studied due to their modifications of BH metrics and their influence on interactions between dark matter and strong gravitational fields \cite{CSR2, CSR5, CSR6}. Analyzing BHs embedded in PFDM environments is crucial for evaluating the effects of dark matter on BH dynamics and for testing theoretical predictions of gravitational models \cite{CSR1, CSR2, CSR6}. Within this framework, optical observables, such as BH shadows and accretion disk imaging, are significantly affected by gravitational lensing near these BH configurations \cite{CSR7, CSR8, CSR9, CSR10, CSR11}. Building on the established formalism of BH thermodynamics in AdS space \cite{CSR13}, investigations in the extended phase space and studies on the role of the cosmological constant on BH potentials have provided quantitative insights \cite{CSR14, CSR15, CSR16, CSR17, CSR18, CSR19}. Also, combining topological classifications, PFDM effects, and analyses of optical and thermodynamic properties yields a systematic approach for studying BHs in astrophysical and cosmological contexts.

The quantum Oppenheimer-Snyder BH model \cite{QOS1, QOS2} establishes a systematic framework for testing gravitational collapse within loop quantum gravity (LQG) \cite{QOS3, QOS4, QOS5, QOS6, QOS7}, extending the classical homogeneous dust collapse formulated by Oppenheimer and Snyder \cite{QOS8}.  Also, within this framework, the inclusion of a Cloud of Strings (CS), representing a continuous distribution of one-dimensional string-like matter across spacetime, introduces an additional source of stress-energy, altering both the internal dynamics and the external geometry.  In this case, the CS generates a repulsive contribution that counterbalances gravitational collapse, thereby modifying the formation of horizons and the causal structure of the resulting BH \cite{QOS9, QOS10}.  In the interior region, originally characterized by a homogeneous isotropic dust distribution, the presence of the string cloud modifies the effective energy density and pressure, which impacts the collapse rate and may delay or prevent singularity formation.  Also, applying the Israel junction conditions \cite{QOS15, QOS16} to connect this modified interior with an exterior spacetime yields an effective BH solution in which the string-induced contributions produce corrections to the metric, particularly in the vicinity of the horizon.  In this case, these corrections result in measurable deviations in quantities such as horizon radii, gravitational lensing, quasi-normal mode (QNMs) spectra, and Hawking radiation, showing that the CS has a significant influence on both the classical and semiclassical properties of BHs \cite{QOS17, QOS18, QOS19, QOS20}. In this context, incorporating the CS provides a consistent mechanism to investigate quantum-corrected collapse, horizon modifications, and potential observational signatures within the LQG formalism.

Our study investigates quantum Oppenheimer-Snyder BHs embedded in a cloud of strings and surrounded by perfect fluid dark matter. We first analyze the spacetime geometry to quantify the modifications induced by quantum corrections, string cloud contributions, and dark matter on the BH’s horizon structure, curvature invariants, and metric functions. Also, the optical properties are then examined through the photon sphere and associated BH shadow, providing criteria for potential observational differentiation from classical BH configurations. In this case, test particle dynamics are evaluated to determine the influence of the surrounding matter on orbital trajectories, stability conditions, and effective potentials. Additionally, scalar field perturbations are employed to characterize the BH models response to external excitations and to estimate all properties. The thermodynamic properties are further tested, illustrating deviations in temperature, entropy, and specific heat ($C_V$) due to string cloud and dark matter effects. In this context, this framework tested quantum corrections and matter distributions within BH physics and shows both theoretical insights and implications for observational constraints on compact objects. 

The paper is organized as follows. In Sec.~\ref{sec:2}, we present the construction of the quantum Oppenheimer-Snyder black hole embedded in a cloud of strings and surrounded by perfect fluid dark matter, detailing the spacetime geometry and the fundamental parameters that define the model. Sec.~\ref{S3} is devoted to the astrophysical signatures of the black hole, focusing on its optical properties, in particular the photon sphere structure and the resulting black hole shadow. In Sec.~\ref{S4}, we study the dynamics of test particles, emphasizing their motion, stability, and effective potentials in the modified gravitational background. Sec.~\ref{S5} examines scalar perturbations, providing insight into wave propagation and associated physical effects. The thermodynamic properties of the black hole are analyzed in Sec.~\ref{S6}. Finally, Sec.~\ref{S7} summarizes our results and discusses possible directions for future research. Throughout the manuscript, we adopt natural units with $c = \hbar = 8 \pi G = 1$.

\section{Spherically Symmetric BH with CS and PFDM }\label{sec:2}

The classical Oppenheimer-Snyder model describes the collapse of a uniformly distributed matter-filled universe into a black hole, ultimately forming a gravitational singularity \cite{JR1939}. By incorporating quantum gravity effects, a quantum-corrected version of the Oppenheimer-Snyder black hole was proposed in Ref.~\cite{JL2023}. The metric of this quantum Oppenheimer--Snyder black hole is given by

\begin{equation}
ds^{2} = -f(r) dt^{2} + f(r)^{-1} dr^{2} + r^{2} d\theta^{2} + r^{2} \sin^{2} \theta d\phi^{2},\label{a1}
\end{equation}

where

\begin{equation} 
f(r) = 1 - \frac{2M}{r} + \frac{\alpha M^{2}}{r^{4}}.\label{a2}
\end{equation}

Here, \(M\) is the Arnowitt--Deser--Misner mass of the black hole, and \(\alpha\) is the parameter characterizing quantum corrections. Equation~\eqref{a1} reduces to the metric of a Schwarzschild black hole when the parameter \(\alpha\) vanishes. For convenience, we define a dimensionless quantum correction parameter \(\hat{\alpha} = \alpha / M^{2}\), which is used throughout this work. Previous studies \cite{LZ2024,HA2025,AV2025} have constrained the parameter \(\hat{\alpha}\) using observations of the black holes' shadows of M87* and Sgr~A*, as well as strong gravitational lensing effects, and found that the upper bound on \(\hat{\alpha}\) is 1.4087.

To construct the Quantum-OP BH coupled with a cloud of strings and PFDM, we need to start with the Nambu-Goto action, which describes strings like objects \cite{PSL1979},   
\begin{equation}
    S^{CS}=\int \sqrt{-\gamma}\,\mathcal{M}\,d\lambda^0\,d\lambda^1=\int \mathcal{M}\sqrt{-\frac{1}{2}\,\Sigma^{\mu \nu}\,\Sigma_{\mu\nu}}\,d\lambda^0\,d\lambda^1,\label{a3}
\end{equation}
where $\mathcal{M}$ is the dimensionless constant which characterizes the string, ($\lambda^0\,\lambda^1$) are the time
like and spacelike coordinate parameters, respectively \cite{JLS1960}. $\gamma$  is the determinant of the induced metric of the strings world sheet given by $\gamma=g^{\mu\nu}\frac{ \partial x^\mu}{\partial \lambda^a}\frac{ \partial x^\nu}{\partial \lambda^b}$.  $\Sigma_{\mu\nu}=\epsilon^{ab}\frac{ \partial x^\mu}{\partial \lambda^a}\frac{ \partial x^\nu}{\partial \lambda^b}$ is bivector related to string world sheet, where $\epsilon^{ ab}$ is the second rank Levi-Civita tensor which takes the non-zero values as $\epsilon^{ 01} = -\epsilon^{ 10} = 1$.

\begin{equation}
   T_{\mu\nu}^{CS}=2 \frac{\partial}{\partial g_{\mu \nu}}\mathcal{M}\sqrt{-\frac{1}{2}\Sigma^{\mu \nu}\,\Sigma_{\mu\nu}} =\frac{\rho^{\rm CoS} \,\Sigma_{\alpha\nu}\, \,\Sigma_{\mu}^\alpha }{\sqrt{-\gamma}}, \label{a4}
 \end{equation}
where $\rho$ is the proper density of the CoS. 

\begin{equation}
    T^{t\,(\rm CS)}_{t}=\rho^{\rm CS}=\frac{\gamma}{r^2}=T^{r\,(\rm CS)}_{r},\quad T^{\theta\,(\rm CS)}_{\theta}=T^{\phi\,(\rm CS)}_{\phi}= 0,\label{a5}
\end{equation}

Incorporating PFDM into the BH provides a natural framework to
examine the interplay between DM and regular BHs. As PFDM captures the large scale DM
distribution influencing galactic and BH environments, its inclusion enables an assessment
of how DM modifies the horizon structure and thermodynamic properties of the Bardeen
BH. Assuming the BH is embedded in PFDM background, the energy momentum tensor is given by \cite{HXZ2021}
\begin{equation}
T^{\mu}_{\ \nu} = \mathrm{diag}(-\rho,\, p_{r},\, p_{\theta},\, p_{\phi}),\label{a6}
\end{equation}
with
\begin{equation}
\rho = -p_{r} = \frac{\lambda}{\kappa r^{3}}, \quad p_{\theta} = p_{\phi} = \frac{\lambda}{2\kappa r^{3}},\label{a7}
\end{equation}

where \(\rho\), \(p_{r}\), and \(p_{\theta} = p_{\phi}\) denote the energy density, radial pressure, and tangential pressures of DM, respectively \cite{VVK2003,MH2012}. The weak energy condition \(T_{tt} \geq 0\) requires \(\lambda \geq 0\) \cite{HXZ2021}. The DM parameter \(\lambda\) characterizes the local DM density and its dynamical effect on the spacetime geometry near the BH, rather than the global cosmological DM content. Observationally, \(\lambda\) can be constrained by comparing theoretical predictions-such as galaxy rotation curves or BH shadow sizes-with astrophysical data. In the following section, we constrain these parameters using the observational data from EHT.

Thereby, the line element describing the Quantum-corrected OS BH with a cloud of string and PFDM is given by
The metric of this quantum Oppenheimer--Snyder black hole is given by

\begin{equation}
ds^{2} = -f(r) dt^{2} + f(r)^{-1} dr^{2} + r^{2} d\theta^{2} + r^{2} \sin^{2} \theta d\phi^{2},\label{metric}
\end{equation}
where
\begin{align}
f(r) &= 1 -\gamma- \frac{2M}{r} + \frac{\alpha M^{2}}{r^{4}}+\frac{\lambda}{r}\ln\!\frac{r}{|\lambda|}\nonumber\\
&=1 -\gamma- \frac{2M}{r} + \frac{\hat{\alpha} M^{4}}{r^{4}}+\frac{\lambda}{r}\ln\!\frac{r}{|\lambda|}.\label{fucntion}
\end{align}

In the limit $\alpha=0$, corresponding to the absence of quantum corrections, the lapse function simplifies as,
\begin{equation}
f(r) = 1 -\gamma- \frac{2M}{r}+\frac{\lambda}{r}\ln\!\frac{r}{|\lambda|}.\label{a8}
\end{equation}
In that case, the space-time (\ref{metric}) is the Letelier black hole surrounded by a perfect fluid dark matter \cite{AS2024}. Moreover,in the limit $\alpha=0=\lambda$, the space-time (\ref{metric}) reduces to the Leterlier black hole \cite{PSL1979}.

\begin{figure}[ht!]
\begin{center}
\begin{tabular}{ccc}
\includegraphics[height=5cm]{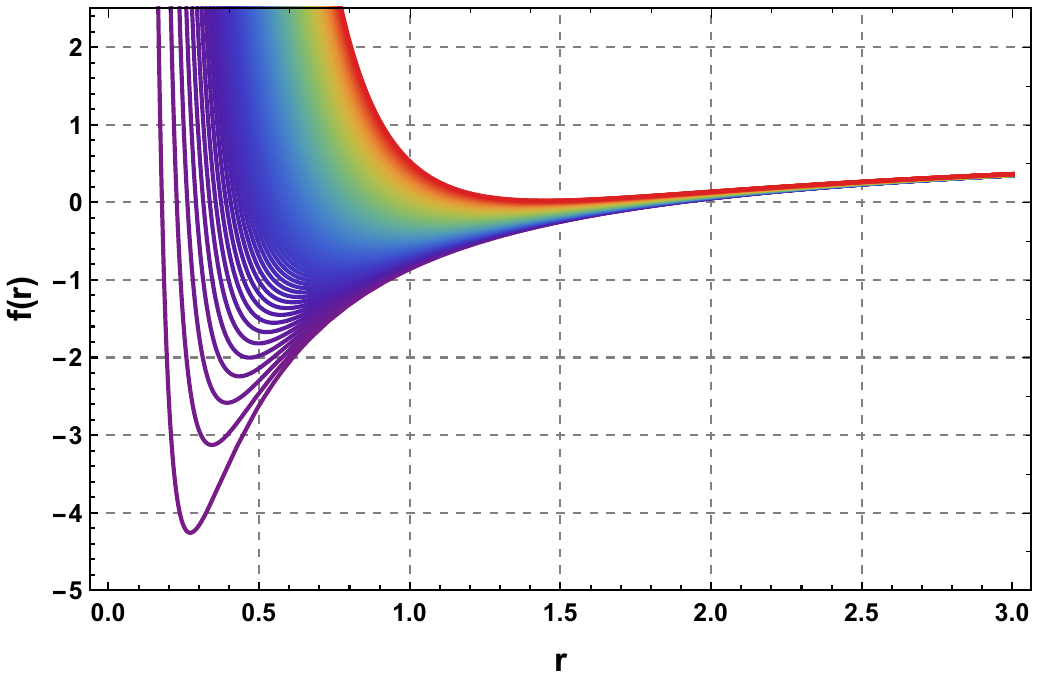} 
\includegraphics[height=5cm]{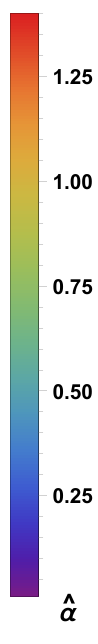}\\
(a) $\lambda=\gamma=0.1$\\
\includegraphics[height=5cm]{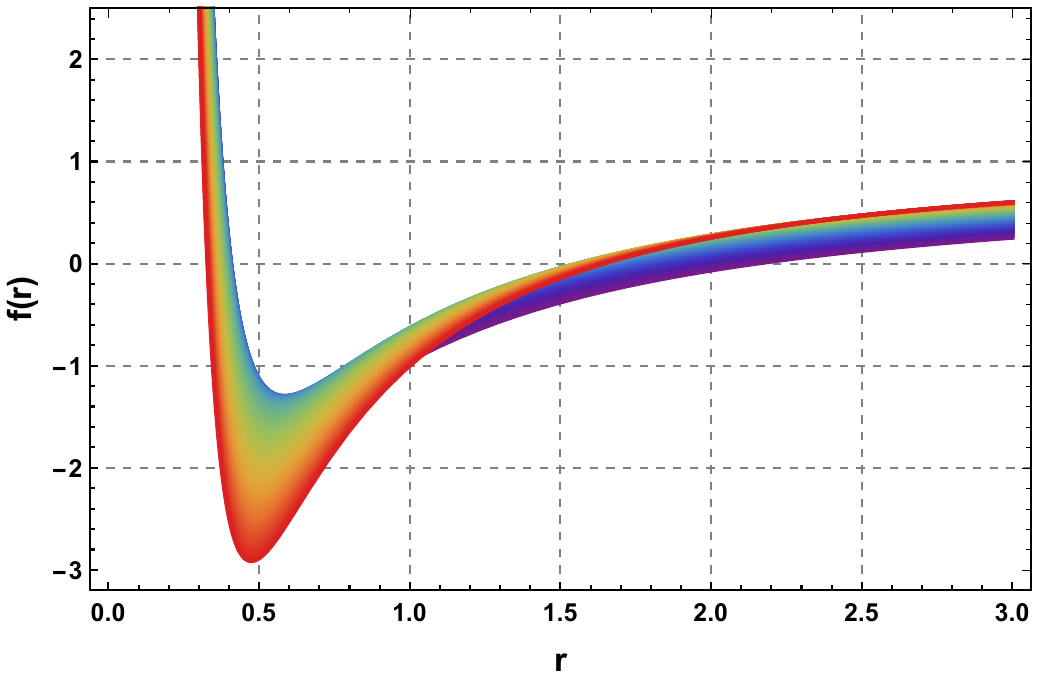} 
\includegraphics[height=5cm]{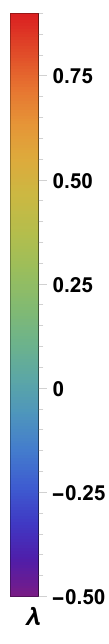}\\
(b) $\hat{\alpha} =\gamma=0.1$\\
\includegraphics[height=5cm]{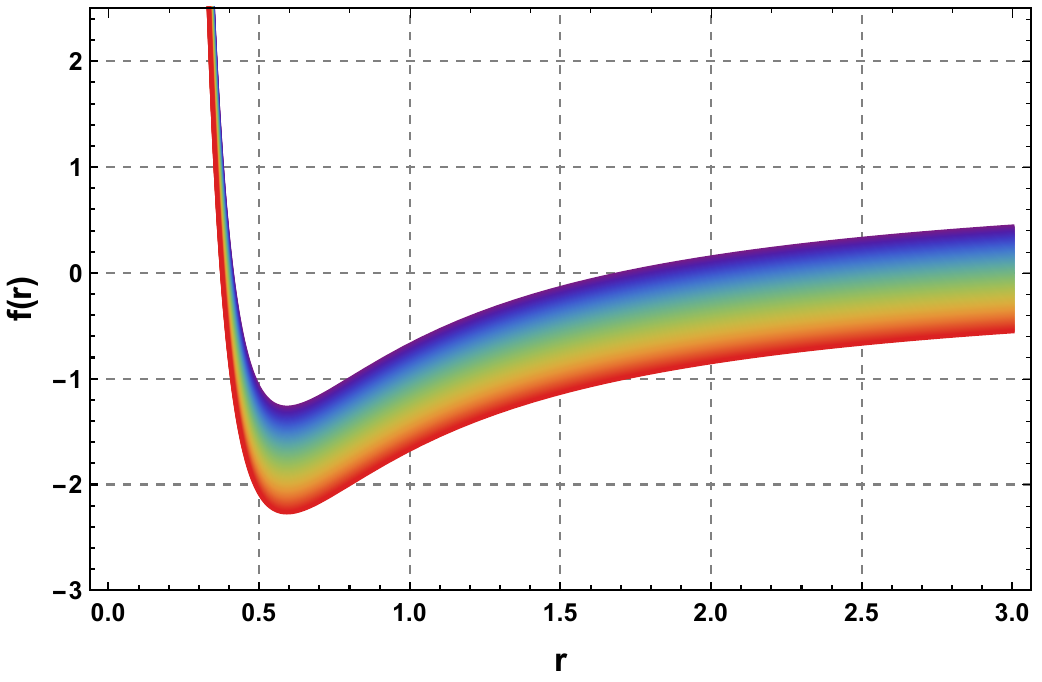} 
\includegraphics[height=5cm]{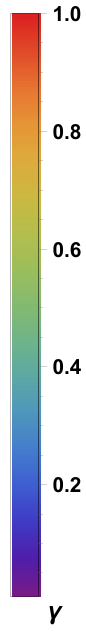}\\
(c) $\lambda=\hat{\alpha} =0.1$
\end{tabular}
\end{center}
\caption{The behavior of the metric function as a function of radial distance by varying $\hat{\alpha},\,\lambda$ and $\gamma$.\label{fig01}}
\end{figure}

Figure~\ref{fig01} illustrates the radial behavior of the lapse function $f(r)$ for different choices of the deformation parameters $\hat{\alpha}$, $\lambda$, and $\gamma$, while keeping the remaining parameters fixed. In Fig.\ref{fig01}(a), variations of the quantum correction parameter $\hat{\alpha}$ significantly modify the near-horizon structure of the spacetime, leading to noticeable shifts in the position and depth of the potential well at small $r$, while the asymptotic behavior remains largely unaffected. Fig.\ref{fig01} (b) shows that the perfect fluid dark matter parameter $\lambda$ introduces a milder deformation, smoothing the metric function close to the origin and slightly shifting the horizon location without altering the qualitative asymptotic flatness. In contrast, Fig.\ref{fig01} (c) demonstrates that the string cloud parameter $\gamma$ induces a global rescaling of the metric function, reducing its asymptotic value and thereby modifying the effective gravitational potential at all radial scales. These results indicate that quantum corrections predominantly influence the strong-field region, whereas the string cloud and PFDM parameters control large-scale deviations from the Schwarzschild geometry, with direct implications for horizon structure, photon dynamics, and thermodynamic stability.

Various scalar curvature are as follows:
\begin{itemize}
    \item Ricci Scalar:\\ 
    \begin{eqnarray}
        R=g^{\mu\nu}R_{\mu\nu}&=&-\,\frac{2}{r^{2}}\left[ f(r)-1 + 2r\,f'(r)\right] - f''(r)\nonumber\\
        &=&\frac{2 r^{4}\gamma - M\left(6M^2\hat{\alpha} + r^{3}\lambda\right)}{r^{6}}.
        \label{a9}
    \end{eqnarray}
    \item The Quadratic Ricci tensor:\\
    \begin{eqnarray}
        R^{\mu\nu}R_{\mu\nu}&=&
-\,\frac{
M^{2}}{2r^{12}}\left[16M^2\hat{\alpha} - 2r^{3}(1+\lambda) + r^{3}\lambda \ln\!\left(\frac{r}{\lambda}\right)\right]\nonumber\\
&\times&
\left[4M^2\hat{\alpha} - r^{3}(2+\lambda) + r^{3}\lambda \ln\!\left(\frac{r}{\lambda}\right)\right]
 .  \label{a11}
    \end{eqnarray}
    \item The Kretschmann scalar:\\
    \begin{eqnarray}
        \mathcal{K}&=&R^{\mu\nu\rho\sigma}R_{\mu\nu\rho\sigma}\nonumber\\ &=&\frac{2\bigl(-1 + f(r)\bigr)^{2}}{r^{4}} + \frac{f'(r)^{2}}{r^{2}}\nonumber\\
&=&\frac{
M^{2}}{r^{12}}\left[4M^2\hat{\alpha}  - r^{3}(2+\lambda) + r^{3}\lambda \ln\left[\frac{r}{\lambda}\right]\right]^{2} \nonumber\\
&+& \frac{2}{r^{12}}\left[M^4\hat{\alpha} -2Mr^{3} - r^{4}\gamma + Mr^{3}\lambda \ln\left(\frac{r}{\lambda}\right)\right]^{2}
 .\label{a11b}\nonumber\\
    \end{eqnarray}
\end{itemize}

\begin{figure}[ht!]
\begin{center}
\begin{tabular}{ccc}
\includegraphics[height=5cm]{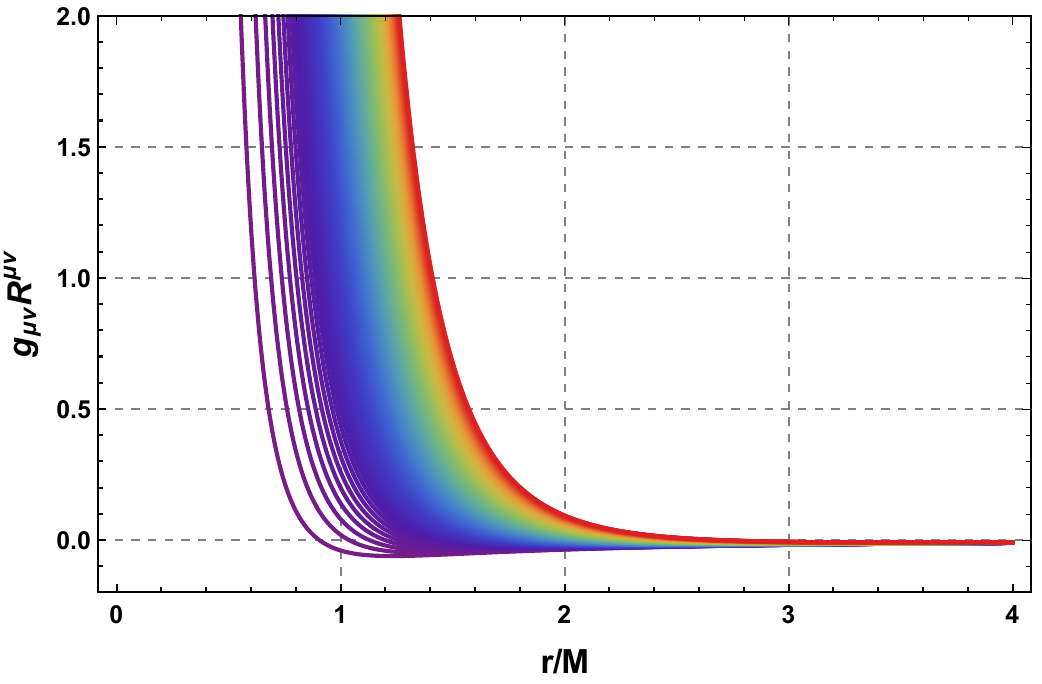} 
\includegraphics[height=5cm]{fig00a.pdf}\\
(a) $\lambda=\gamma=0.1$\\
\includegraphics[height=5cm]{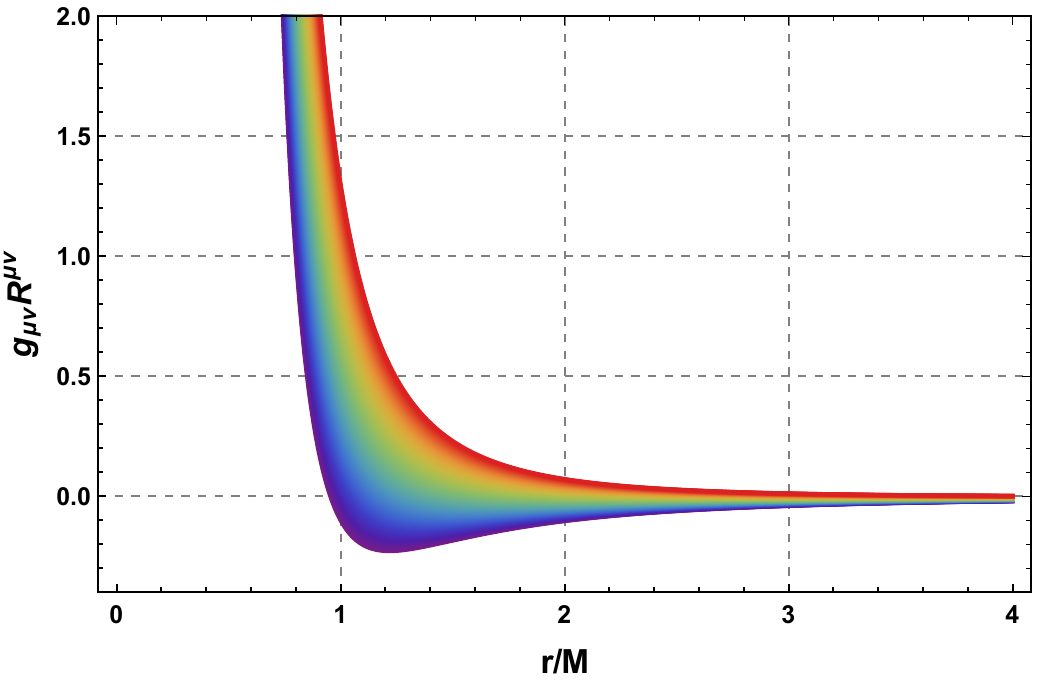} 
\includegraphics[height=5cm]{fig00b.pdf}\\
(b) $\hat{\alpha} =\gamma=0.1$\\
\includegraphics[height=5cm]{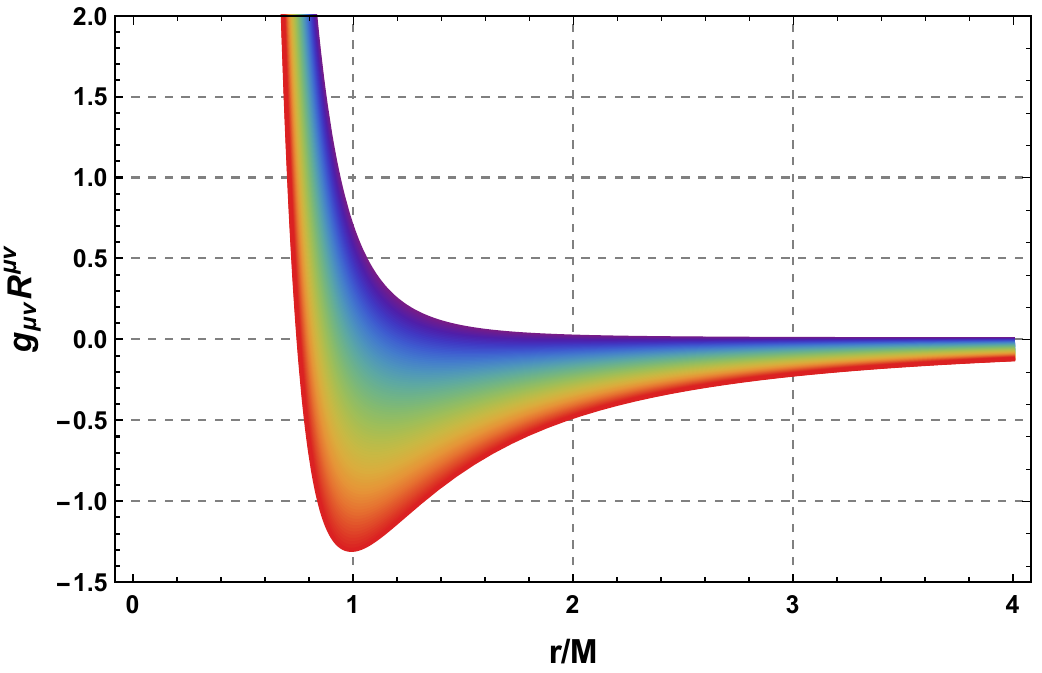} 
\includegraphics[height=5cm]{fig00c.pdf}\\
(c) $\lambda=\hat{\alpha} =0.1$
\end{tabular}
\end{center}
\caption{The behavior of the Ricci scalar as a function of radial distance by varying $\hat{\alpha},\,\lambda$ and $\gamma$.
\label{fig00}}
\end{figure}

Figure~\ref{fig00} displays the radial profile of the Ricci scalar $R$ for different values of the deformation parameters $\hat{\alpha}$, $\lambda$, and $\gamma$, while keeping the remaining parameters fixed. In Fig.\ref{fig00}(a), increasing the quantum correction parameter $\hat{\alpha}$ markedly alters the curvature behavior in the strong-field regime, leading to pronounced variations of $R$ near the central region and signaling the sensitivity of spacetime curvature to higher-order quantum corrections. Fig.\ref{fig00}(b) shows that the perfect fluid dark matter parameter $\lambda$ primarily affects the intermediate radial domain, smoothing the curvature profile and reducing the magnitude of the Ricci scalar without qualitatively changing its asymptotic behavior. In Fig.\ref{fig00}(c), variations of the string cloud parameter $\gamma$ induce a global shift in the Ricci scalar, enhancing curvature effects close to the origin and modifying the large-scale curvature structure of the spacetime. In all cases, the Ricci scalar approaches zero at large radial distances, consistently indicating asymptotic flatness. These results demonstrate that the combined effects of quantum corrections, string clouds, and PFDM significantly reshape the curvature structure of the black hole spacetime, with potential consequences for horizon formation, geodesic motion, and the stability of perturbative fields.

\begin{figure}[ht!]
\begin{center}
\begin{tabular}{ccc}
\includegraphics[height=5cm]{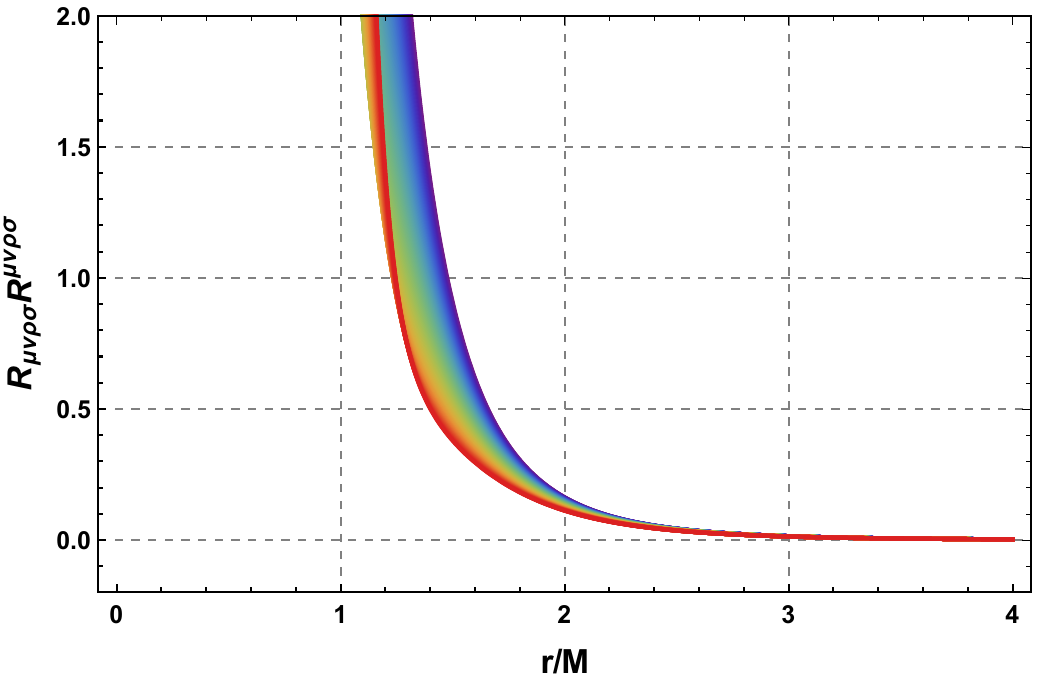} 
\includegraphics[height=5cm]{fig00a.pdf}\\
(a) $\lambda=\gamma=0.1$\\
\includegraphics[height=5cm]{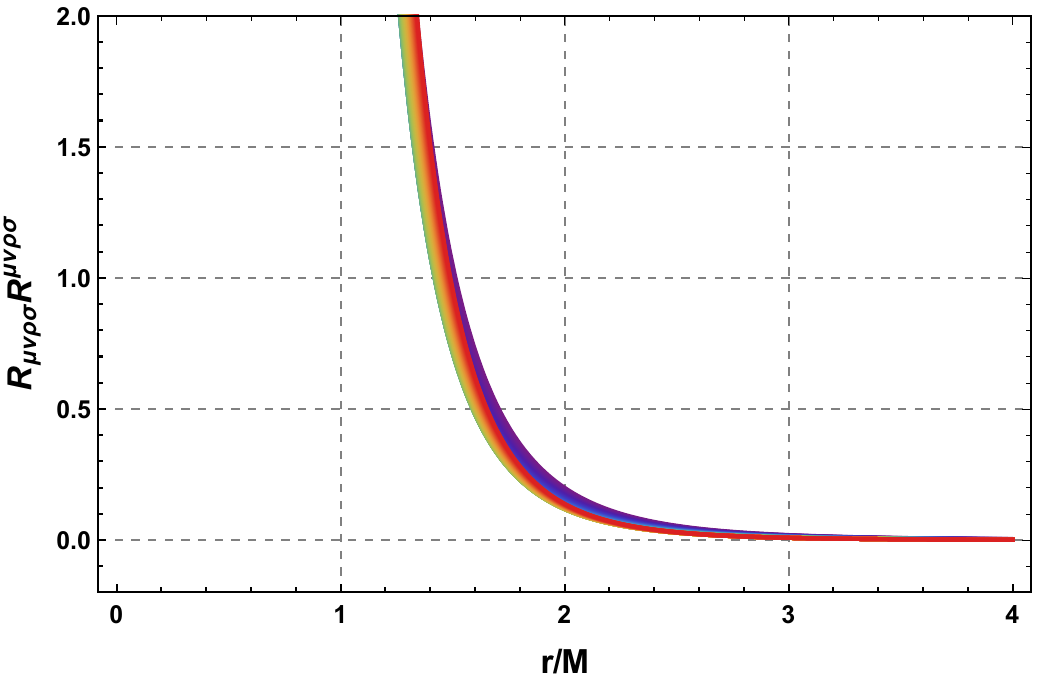} 
\includegraphics[height=5cm]{fig00b.pdf}\\
(b) $\hat{\alpha} =\gamma=0.1$\\
\includegraphics[height=5cm]{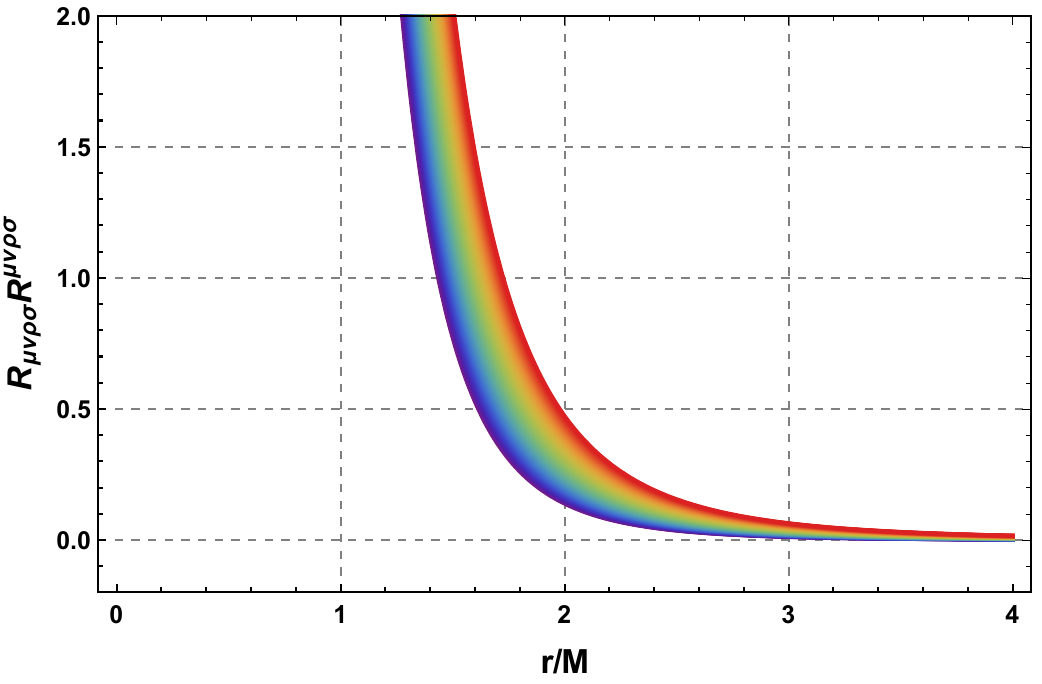} 
\includegraphics[height=5cm]{fig00c.pdf}\\
(c) $\lambda=\hat{\alpha} =0.1$
\end{tabular}
\end{center}
\caption{The behavior of the Kretschmann scalar as a function of radial distance by varying $\hat{\alpha},\,\lambda$ and $\gamma$.
\label{fig02}}
\end{figure}

Figure~\ref{fig02} presents the radial behavior of the Kretschmann scalar $\mathcal{K}=R_{\mu\nu\rho\sigma}R^{\mu\nu\rho\sigma}$ for different values of the deformation parameters $\hat{\alpha}$, $\lambda$, and $\gamma$, with the remaining parameters held fixed. In Fig.\ref{fig02} (a) and (b), variations of the quantum correction parameter $\hat{\alpha}$ and the perfect fluid dark matter parameter $\lambda$ primarily affect the magnitude of the curvature invariant in the strong-field region, leading to noticeable quantitative changes near the central region while preserving the overall monotonic decay of $\mathcal{K}$ with increasing radius. Fig.\ref{fig02} (c) shows that the string cloud parameter $\gamma$ induces a more pronounced global modification, enhancing the curvature scale at small and intermediate radii and delaying the decay of $\mathcal{K}$ toward its asymptotic value. In all cases, the Kretschmann scalar decreases rapidly and approaches zero at large distances, confirming the asymptotic flatness of the spacetime. The divergence of $\mathcal{K}$ as $r \to 0$ indicates the persistence of a curvature singularity at the origin, although its strength and radial profile are significantly influenced by the combined effects of quantum corrections, string clouds, and perfect fluid dark matter. These features play a crucial role in determining the near-horizon geometry and may impact the stability and observational signatures of the black hole.

The existence of an event horizon in the present black hole spacetime is closely connected to the behavior of curvature invariants in the vicinity of the strong-field region, as illustrated in Figs.~\ref{fig00} and~\ref{fig02}. Although both the Ricci scalar and the Kretschmann scalar diverge as $r \to 0$, signaling the presence of a central curvature singularity, they remain finite and well behaved at the horizon radius $r=r_{h}$, which is defined by the largest positive root of the lapse function $f(r)=0$. This regular behavior of curvature invariants at $r=r_{h}$ indicates that the singularity is hidden behind an event horizon, thereby satisfying the cosmic censorship conjecture within the parameter ranges considered. Moreover, the smooth radial profiles of $R$ and $\mathcal{K}$ outside the horizon confirm that the spacetime geometry remains physically admissible in the exterior region, ensuring the causal separation between the singular core and distant observers. Consequently, despite the presence of quantum corrections, a string cloud, and perfect fluid dark matter, the spacetime retains a genuine black hole character rather than describing a naked singularity, with the event horizon playing a crucial role in shielding the curvature divergences from the external universe.

\section{Astrophysical Signature of BH}\label{S3}

Geodesic motion plays a central role in black hole physics, as it governs the trajectories of matter and light in the curved spacetime around a black hole. The properties of geodesics encode the structure of the black hole's gravitational field and provide the basis for understanding a wide range of astrophysical and observational phenomena. Recent observations by the Event Horizon Telescope (EHT) have further highlighted the importance of geodesic studies in exploring black hole physics. For related studies, see, for example, \cite{FA1,FA2,FA3,FA4,FA5,FA6} and the references therein.

In the current study, we study geodesic motion through the Lagrangian density function as follows:
\begin{equation}
    \mathbb{L}=\frac{1}{2}\,g_{\mu\nu}\,\dot x^{\mu}\,\dot x^{\nu},\label{aa1}
\end{equation}
where dot represents derivative w.r to an affine parameter $\tau$. For null geodesics, we have $2\mathbb{L}=0$.

Using the given metric (\ref{metric}), the explicit form of the Lagrangian density function is given by 
\begin{align}
    \mathbb{L}=\frac{1}{2}\,\Big[-f(r)\,\dot t^{2}+\frac{1}{f(r)}\,\dot r^2+r^{2}\,\dot \theta^2+r^2\,\sin^{2}\theta\,\dot \phi^{2}\Big].\label{aa2}
\end{align}

Considering the geodesic motion in the equatorial plane, defined by $\theta=\pi/2$ and $\frac{d\theta}{d\tau}=0$. From the above expression, we observed that the Lagrangian density function $\mathbb{L}=\mathbb{L}(r)$ depends on the radial coordinate, while it is independent of the temporal $t$ and the azimuthal coordinates $\phi$. Therefore, there exist two constants of motion corresponding to these cyclic coordinates. The conserved quantity associated with the temporal coordinate is called the conserved energy $\mathrm{E}$ and the conserved quantity associated with the azimuthal coordinate $\phi$ is the conserved angular momentum $\mathrm{L}$. There are given by
\begin{equation}
    \mathrm{E}=f(r)\,\dot t,\qquad \mathrm{L}=r^2\,\dot \phi.\label{aa3}
\end{equation}

With these, the motion of photon particles satisfying the condition $g_{\mu\nu}\,\dot x^{\mu}\,\dot x^{\nu}=0$ on the equatorial plane is obtain after simplification as,
\begin{equation}
    \dot r^2+V_\text{eff}(r)=\mathrm{E}^2,\label{aa4}
\end{equation}
which is equivalent to one-dimensional equation of motion of particles with the effective potential $V_{\rm eff}$. This potential is of the form 
\begin{equation}
    V_\text{eff}(r)=f(r)\,\frac{\mathrm{L}^2}{r^2}=\frac{\mathrm{L}^2}{r^2}\left(1 -\gamma- \frac{2M}{r} + \frac{\hat{\alpha} M^{4}}{r^{4}}+\frac{\lambda}{r}\ln\!\frac{r}{|\lambda|}\right).\label{aa5}
\end{equation}

From the expression in Eq.~(\ref{aa5}), it is evident that the geometric parameters, such as the black hole mass $M$, the string cloud parameter $\gamma$, the quantum correction parameter $\hat{\alpha}$, and the perfect fluid parameter $\lambda$, significantly modify the effective potential governing photon dynamics. Variations in these parameters alter the spacetime curvature and, consequently, affect observable quantities such as the photon sphere and the black hole shadow.

\begin{figure}[ht!]
\begin{center}
\begin{tabular}{ccc}
\includegraphics[height=5cm]{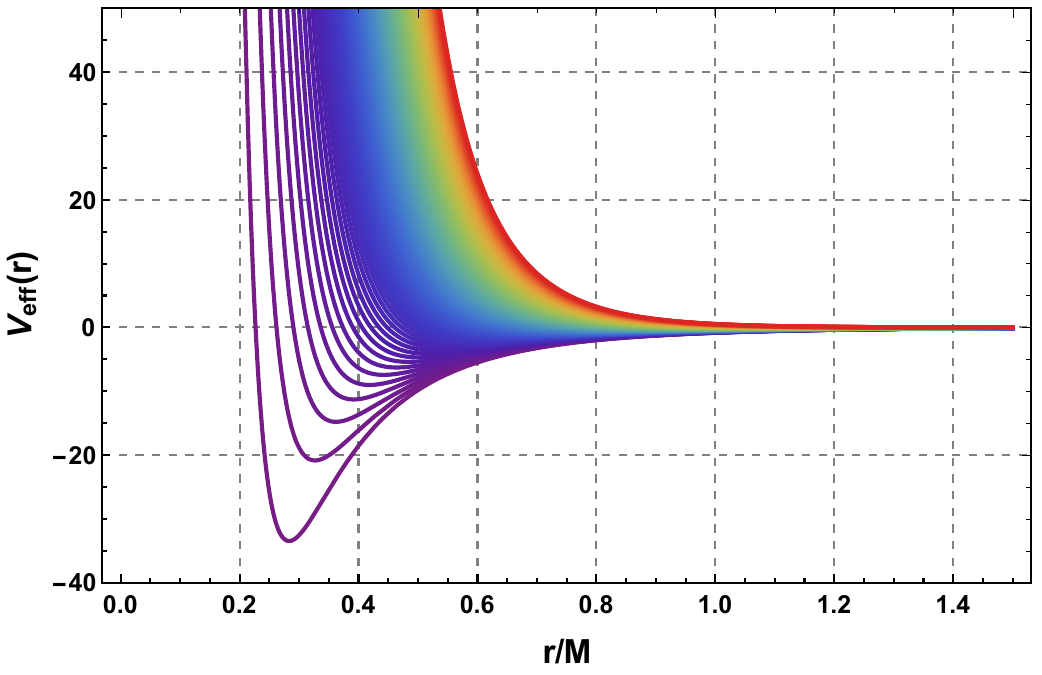} 
\includegraphics[height=5cm]{fig00a.pdf}\\
(a) $\lambda=\gamma=0.1$\\
\includegraphics[height=5cm]{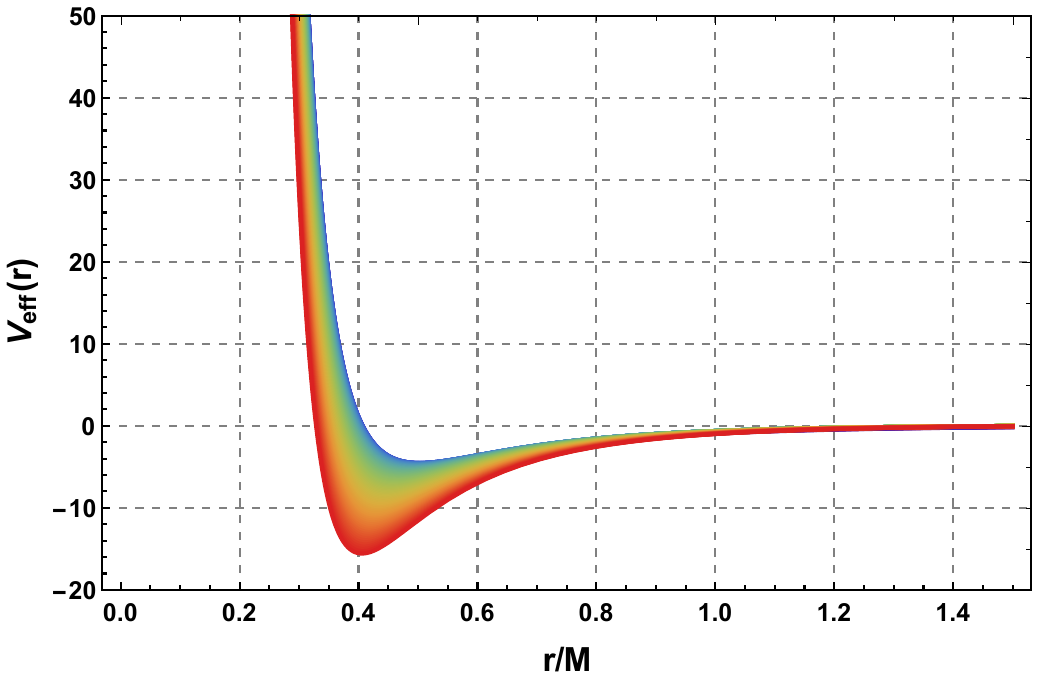} 
\includegraphics[height=5cm]{fig00b.pdf}\\
(b) $\hat{\alpha}=\gamma=0.1$\\
\includegraphics[height=5cm]{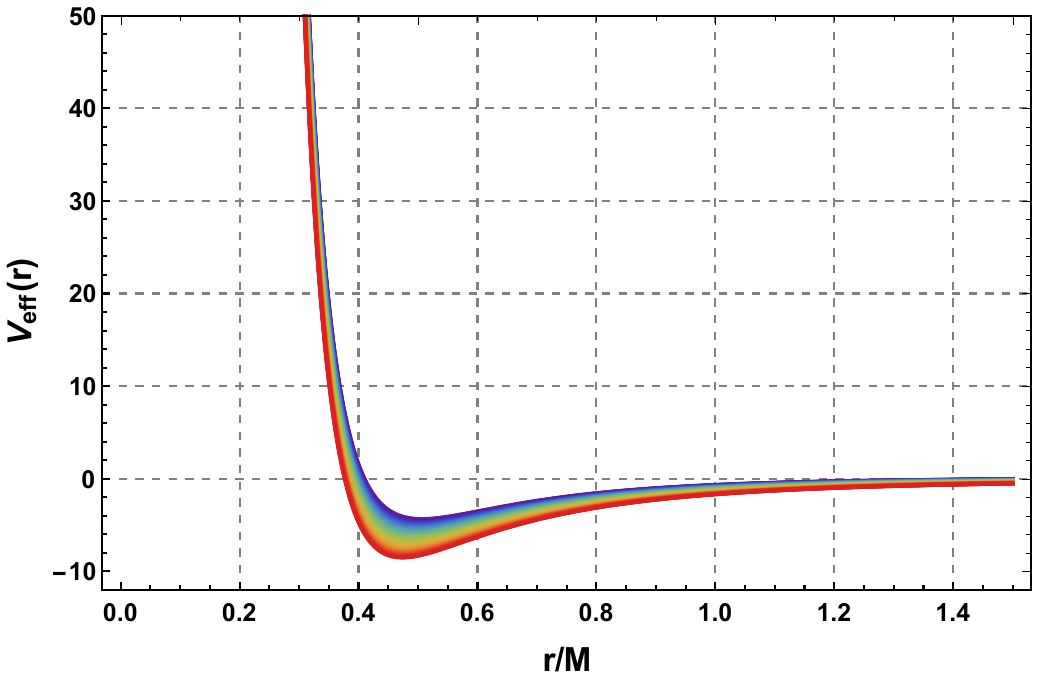} 
\includegraphics[height=5cm]{fig00c.pdf}\\
(c) $\lambda=\hat{\alpha}=0.1$
\end{tabular}
\end{center}
\caption{The behavior of the effective potential governing the photon dynamics as a function of the radial distance. Here, the conserved angular momentum is set to $L=1$, and the black hole mass is chosen to be unity for simplicity.
\label{fig03}}
\end{figure}

Figure~\ref{fig03} depicts the radial behavior of the effective potential $V_{\rm eff}(r)$ for photon motion under variations of the deformation parameters $\hat{\alpha}$, $\lambda$, and $\gamma$, while keeping the remaining parameters fixed. In Fig.\ref{fig03}(a), where $\lambda=\gamma=0.1$, increasing the quantum correction parameter $\hat{\alpha}$ produces significant modifications in the strong-field region, deepening the potential well and shifting the position of its maximum, which directly affects the location of unstable photon orbits. Fig.\ref{fig03}(b) illustrates the influence of the perfect fluid dark matter parameter $\lambda$ for $\hat{\alpha}=\gamma=0.1$, showing that PFDM smooths the effective potential near the horizon and reduces the steepness of the potential barrier, indicating a milder but non-negligible impact on photon dynamics. In contrast, Fig.\ref{fig03}(c) demonstrates that the string cloud parameter $\gamma$, for $\lambda=\hat{\alpha}=0.1$, induces a global downward shift of the effective potential, reflecting a deficit in the asymptotic structure of the spacetime and altering photon trajectories at all radial scales. Overall, these results reveal that quantum corrections predominantly govern the near-horizon behavior, while the string cloud and PFDM parameters control large-scale deviations from the Schwarzschild potential, with direct consequences for the photon sphere, gravitational lensing, and the observable size of the black hole shadow.

\subsection{Photon Sphere}\label{S3-1}

The photon sphere is a spherical region around a black hole where photons can travel on unstable circular null geodesics. Its radius depends sensitively on the underlying space-time geometry and therefore on the black hole parameters. Although the photon sphere itself is not directly observable, it plays a central role in shaping the black hole shadow. Photons with impact parameters close to the critical value associated with the photon sphere undergo multiple orbits before escaping to infinity, leading to strong gravitational lensing. This effect produces the bright, narrow emission ring-often referred to as the photon ring-surrounding the dark shadow region in EHT images.

For circular orbits, the conditions $\dot r=0$ and $\ddot r=0$ must be satisfied. Using Eq.~(\ref{aa4}), we find the following relations:
\begin{equation}
\mathrm{E}^2=\frac{\mathrm{L}^2}{r^2}\left(1 -\gamma- \frac{2M}{r} + \frac{\hat{\alpha} M^{4}}{r^{4}}+\frac{\lambda}{r}\ln\!\frac{r}{|\lambda|}\right).\label{aa6}
\end{equation}
And
\begin{equation}
    \frac{d}{dr}\left(\frac{f(r)}{r^2}\right)=0.\label{aa7}
\end{equation}

The photon sphere radius $r=r_{\rm ph}$ can be determined using (\ref{aa7}) and is given by the following polynomial equation:
\begin{equation}
1 -\gamma- \frac{3M}{r_{\rm ph}} + \frac{3\hat{\alpha} M^{4}}{r^{4}_{\rm ph}}+\frac{\lambda}{2r_{\rm ph}}\left(3\ln\!\frac{r_{\rm ph}}{|\lambda|}-1\right)=0.\label{aa8}
\end{equation}
The above polynomial can't be solved analytically due to the presence of the logarithmic function. However, for suitable values of the geometric parameters ($\gamma,\,\lambda,\,\alpha$), one can find numerical values of this radius $r_{\rm ph}$.

\begin{table}[h!]
\centering
\caption{Photon sphere radius $r_{\rm ph}/M$ for different values of $\gamma$ and $\lambda/M$ (with $\hat{\alpha}=1.4$).}
\begin{tabular}{|c| c| c| c| c| c|}
\hline
$\lambda/M (\downarrow) \backslash \gamma (\rightarrow) $ & $0.00$ & $0.05$ & $0.10$ & $0.15$ & $0.20$ \\
\hline

-0.4 & 4.14370 & 4.41209 & 4.71052 & 5.04480 & 5.42215 \\
\hline
-0.2 & 3.69109 & 3.91870 & 4.17036 & 4.45072 & 4.76556 \\
\hline
 0.2 & 1.83125 & 1.94060 & 2.21228 & 2.41784 & 2.61797 \\
 \hline
 0.4 & 1.79624 & 1.79623 & 1.79624 & 2.02459 & 2.24115 \\
 \hline
 0.6 & 1.72956 & 1.72955 & 1.72955 & 1.98287 & 2.16375 \\
\hline
\end{tabular}
\end{table}

\begin{figure}
    \centering
    \includegraphics[width=1\linewidth]{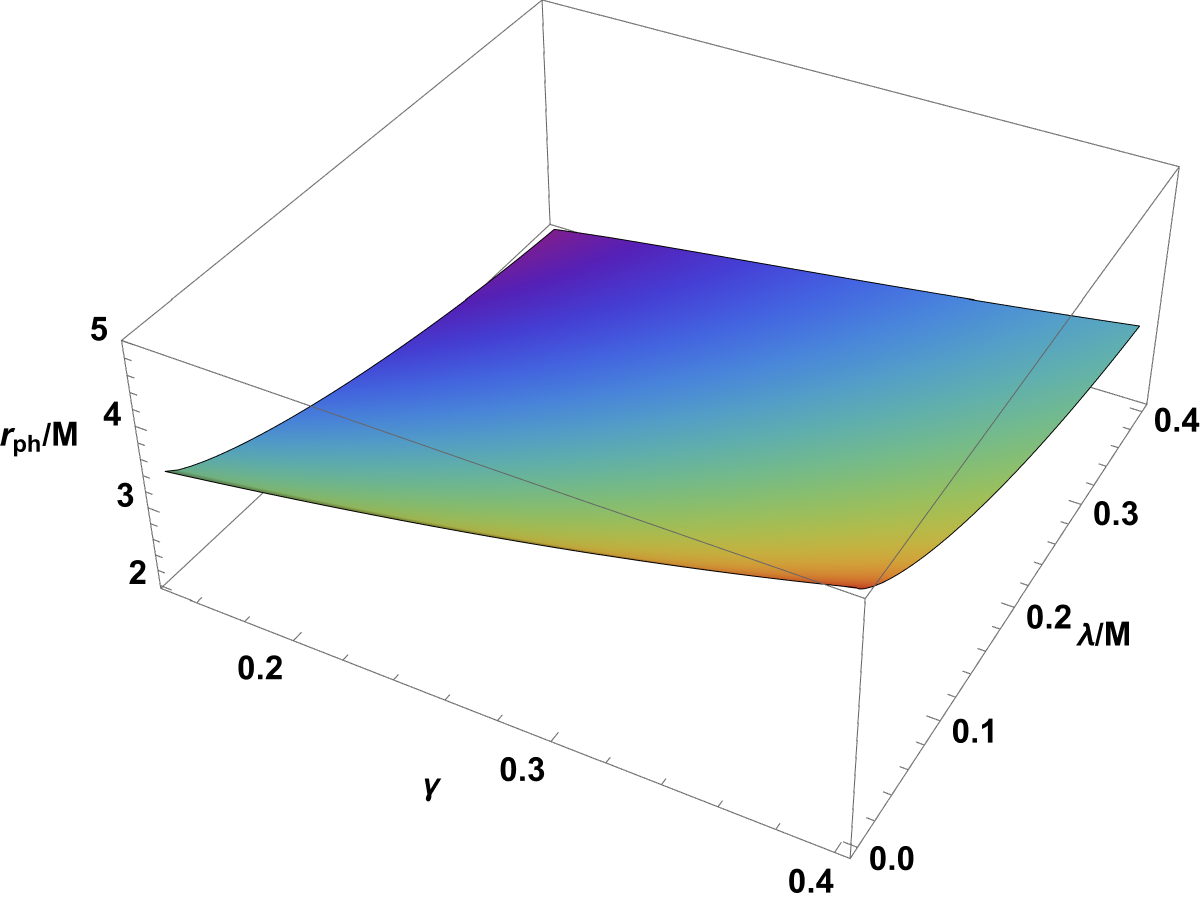}
    \caption{Photon sphere radius with $\hat{\alpha}=1.4$.}
    \label{fig:photon}
\end{figure}

\subsection{BH Shadow}\label{S3-2}

The black hole shadow corresponds to the set of photon trajectories that are captured by the black hole and therefore fail to reach a distant observer. It is a purely gravitational effect determined by the structure of null geodesics in the strong-field region and should not be identified with the event horizon itself. The boundary of the shadow is set by the critical impact parameter associated with the unstable photon sphere, which separates plunging photon orbits from those that escape to infinity. Photons with impact parameters close to this critical value experience extreme gravitational lensing, giving rise to a bright photon ring surrounding the shadow. The size and morphology of the black hole shadow depend sensitively on the underlying spacetime geometry and black hole parameters such as mass and spin, as well as on possible deviations from the Kerr solution. The first horizon-scale images of the supermassive black holes \(M87^{*}\) \cite{EHTL1,EHTL4,EHTL6} and \(Sgr A^{*}\) \cite{EHTL12,EHTL14,EHTL15,EHTL16,EHTL17} obtained by the Event Horizon Telescope (EHT) has gained significant interest on black holes models in general relativity as well as modified gravity theories.

Following the procedure and methodology outlined in \cite{Volker2022}, we determine the shadow radius for a static observer located at position $r_O$ from the center of black hole. Accordingly, the shadow radius is 
\begin{equation}
    R_{\rm sh}=r_{\rm ph}\sqrt{\frac{f(r_O)}{f(r_{\rm ph})}}=r_{\rm ph}\sqrt{\frac{1 -\gamma- \frac{2M}{r_{O}} + \frac{\hat{\alpha} M^{4}}{r^{4}_{O}}+\frac{\lambda}{r_{O}}\,\ln\!\frac{r_{O}}{|\lambda|}}{1 -\gamma- \frac{2M}{r_{\rm ph}} + \frac{\hat{\alpha} M^{4}}{r^{4}_{\rm ph}}+\frac{\lambda}{r_{\rm ph}}\,\ln\!\frac{r_{\rm ph}}{|\lambda|}}}.\label{aa10}
\end{equation}
For a distant observer, the shadow radius simplifies as
\begin{align}
R_{\rm sh}&=r_{\rm ph}\sqrt{\frac{f(r_O \to \infty)}{f(r_{\rm ph})}}\nonumber\\
&=\frac{r_{\rm ph}\,(1-\gamma)^{1/2}}{\sqrt{1 -\gamma- \frac{2M}{r_{\rm ph}} + \frac{\hat{\alpha} M^{4}}{r^{4}_{\rm ph}}+\frac{\lambda}{r_{\rm ph}}\,\ln\!\frac{r_{\rm ph}}{|\lambda|}}}.\label{aa11}
\end{align}
From the expressions (\ref{aa9}) and (\ref{aa11}), we observe that the shadow radius for a static distant observe is equal to $(1-\gamma)^{1/2}$ times the critical impact parameter for photon particles.

\begin{table}[h!]
\centering
\caption{Shadow radius $R_{\rm sh}/M$ for different values of $\gamma$ and $\lambda/M$ (with $\hat{\alpha}=1.4$, $r_O/M=50$).}
\begin{tabular}{|c| c| c| c| c| c|}
\hline
$\lambda/M (\downarrow) \backslash \gamma (\rightarrow) $ & $0.00$ & $0.05$ & $0.10$ & $0.15$ & $0.20$ \\
\hline

-0.6 & 7.95263 & 8.45656 & 9.01692 & 9.64381 & 10.34970 \\
\hline
-0.4 & 7.30569 & 7.74539 & 8.23348 & 8.77861 & 9.39147 \\
\hline
-0.2 & 6.44394 & 6.81338 & 7.22223 & 7.67754 & 8.18803 \\
\hline
 0.2 & 3.46574 & 3.72993 & 3.98622 & 4.24499 & 4.52032 \\
 \hline
 0.4 & 3.01046 & 3.16513 & 3.36854 & 3.61714 & 3.84970 \\

\hline
\end{tabular}
\end{table}

\begin{figure}
    \centering
    \includegraphics[width=1\linewidth]{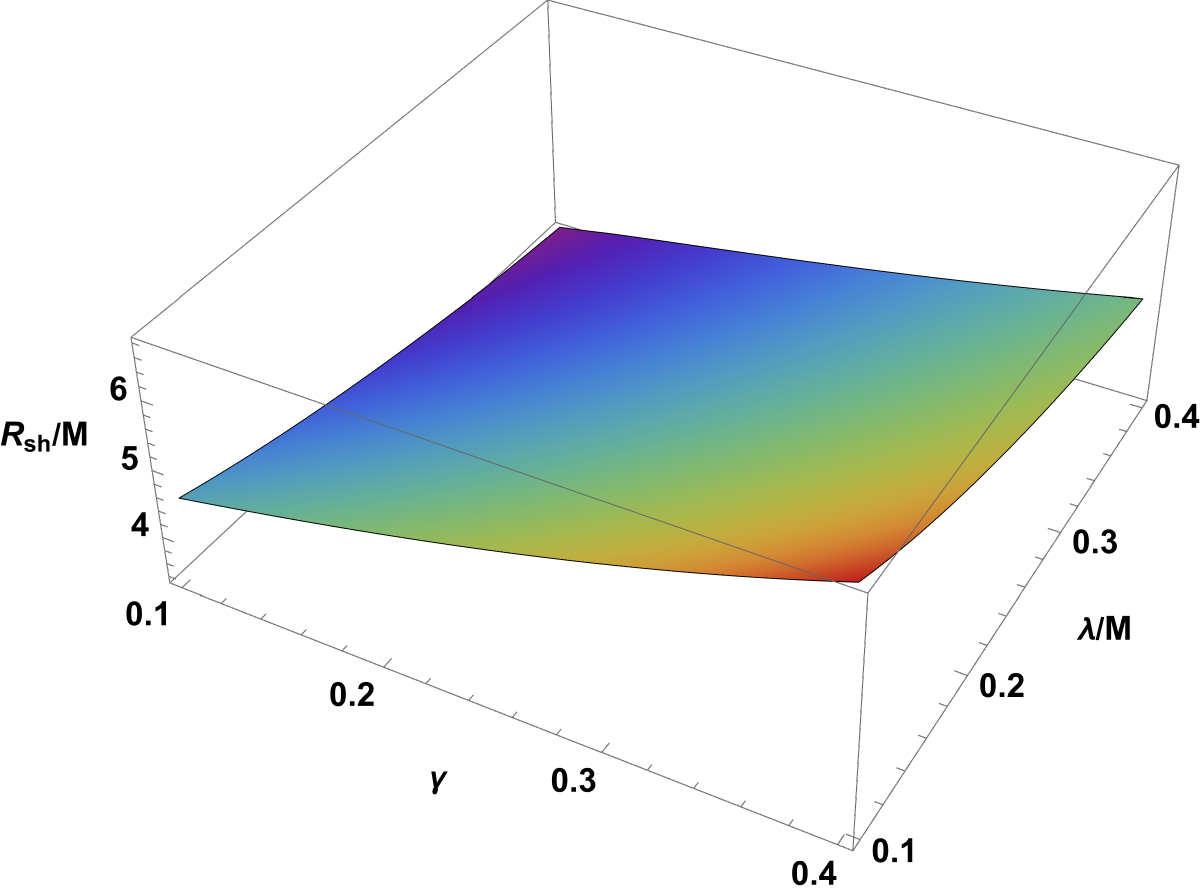}
    \caption{Shadow radius with $\hat{\alpha}=1.4$.}
    \label{fig:shadow}
\end{figure}

\subsection{Photon Trajectories}

In this part, we study photon trajectories showing how the matter field PFDM, string cloud and deformation parameters deviate the photons paths.

At radius $r=r_{\rm ph}$, the critical impact parameter for photons is given by
\begin{equation}
    \beta_{\rm ph}=\frac{\mathrm{E}}{\mathrm{L}}\Big{|}_{r=r_{\rm ph}}=\frac{r_{\rm ph}}{\sqrt{1 -\gamma- \frac{2M}{r_{\rm ph}} + \frac{\hat{\alpha} M^{4}}{r^{4}_{\rm ph}}+\frac{\lambda}{r_{\rm ph}}\,\ln\!\frac{r_{\rm ph}}{|\lambda|}}}.\label{aa9}
\end{equation}
The impact parameter $\beta = \mathrm{L}/\mathrm{E}$ characterizes the trajectory of a photon as observed from far distance and determines whether the photon is captured by the black hole or escapes away to infinity. Photons with $\beta < \beta_{\rm ph}$ inevitably fall into the black hole, while those with $\beta > \beta_{\rm ph}$ are scattered back to infinity after experiencing strong gravitational lensing. The critical impact parameter marks the boundary between capture and escape trajectories and directly determines the apparent size of the black hole shadow. As a result, modification to the spacetime geometry by the string cloud and perfect fluid shifts the photon sphere radius leads to a corresponding change in $\beta_{\rm ph}$, making the critical impact parameter a key observable in strong-field gravity tests and in interpreting Event Horizon Telescope observations.

The equation of orbit is given by
\begin{equation}
    \left(\frac{dr}{d\phi}\right)^2=r^4\left[\frac{1}{\beta^2}-\frac{1}{r^2}\,\left(1 -\gamma- \frac{2M}{r} + \frac{\hat{\alpha} M^{4}}{r^{4}}+\frac{\lambda}{r}\ln\!\frac{r}{|\lambda|}\right)\right].\label{zz1}
\end{equation}
Transforming to a new variable via $r(\phi)=u(\phi)$ and after simplification results
\begin{equation}
\left(\frac{du}{d\phi}\right)^2+(1-\gamma)\,u^2=\frac{1}{\beta^2}+2 M u^3-\hat{\alpha} M^{4}u^{6}+\lambda\,u^3\ln\!(u\,|\lambda|).\label{zz2}
\end{equation}

\begin{itemize}
    \item When $\lambda=0$, the photon trajectories simplifies as,
    \begin{equation}
\left(\frac{du}{d\phi}\right)^2+(1-\gamma)\,u^2=\frac{1}{\beta^2}+2 M u^3-\hat{\alpha} M^{4}u^{6}.\label{zz3}
\end{equation}
Differentiating both sides w. r. to $\phi$ and after simplification results
\begin{equation}
\frac{d^2u}{d\phi^2}+(1-\gamma)\,u=3 M u^2-3\hat{\alpha} M^{4}u^{5}.\label{zz4}
\end{equation}
One can see that the photon trajectories still get modifications by the string cloud as well as the deformation parameters in comparison to the standard result for Schwrazschild BH case. 
\end{itemize}

\section{Test Particle Dynamics around BH }\label{S4}

One of the key applications of geodesic motion is in modeling accretion disks around black holes. The inner edge of such disks typically lies at the innermost stable circular orbit (ISCO), whose location depends on the black hole's mass and spin. This orbit sets important physical limits on the disk's structure and influences the emitted radiation spectrum, making it essential for interpreting observations from X-ray binaries and active galactic nuclei. Similarly, the trajectories of light rays-null geodesics-determine the appearance of black hole shadows, such as those imaged by the Event Horizon Telescope. The shape and size of the shadow are dictated by the geometry of spacetime, particularly by the photon sphere where light can orbit the black hole. For a detailed discussion and related studies, see, for example, \cite{FA1,FA2,FA3,FA4,FA5,FA6,FA7,FA8} and the references therein.

The motion of test particle can be determined by the Lagrangian formalism 
\begin{equation}
    \mathbb{L}=\frac{1}{2} m g_{\mu\nu} \frac{dx^{\mu}}{d\zeta}\,\frac{dx^{\nu}}{d\zeta},\label{dd3}
\end{equation}
In our case at hand, we find explicit form of the Lagrangian density function as
\begin{align}
\mathbb{L}&=\frac{1}{2} m \Big[-f(r) \left(\frac{dt}{d\zeta}\right)^2+\frac{1}{f(r)} \left(\frac{dr}{d\zeta}\right)^2+r^2\left(\frac{d\theta}{d\zeta}\right)^2\nonumber\\
&+r^2 \sin^2 \theta \left(\frac{d\phi}{d\zeta}\right)^2\Big].\label{dd4}
\end{align}

The conserved quantities per unit mass reduce the equations for $t$ and $\phi$ into first integrals:
\begin{align}
    &\frac{p_t}{m}=-\mathcal{E}=-\frac{dt}{d\zeta}\,f(r)\Rightarrow \frac{dt}{d\zeta}=\frac{\mathcal{E}}{f(r)},\label{dd5}\\
    &\frac{p_{\phi}}{m}=\mathcal{L}=r^2 \sin^2 \theta \frac{d\phi}{d\zeta}\Rightarrow \frac{d\phi}{d\zeta}=\frac{\mathcal{L}}{r^2 \sin^2 \theta}.\label{dd6}
\end{align}
Moreover, the angular momentum $p_{\theta}$ using the Euler-Lagrange formula is given by
\begin{equation}
    \frac{p_{\theta}}{m}=r^2 \frac{d\theta}{d\zeta}.\label{dd7}
\end{equation}

Using Eqs. (\ref{dd5}) and (\ref{dd6}), we find the radial component motion of massive test particles on the equatorial plane ($\theta=\pi/2$) as,
\begin{equation}
    \left(\frac{dr}{d\zeta}\right)^2=\mathcal{E}^2-\mathbb{U}_{\rm eff} (r),\label{dd8}
\end{equation}
where the effective potential of the system is given by
\begin{align}
\mathbb{U}_{\rm eff} (r)&=\left(1+\frac{\mathcal{L}^2}{r^2}\right)\,f(r)\nonumber\\
&=\left(1+\frac{\mathcal{L}^2}{r^2}\right)\left(1 -\gamma- \frac{2M}{r} + \frac{\hat{\alpha} M^{4}}{r^{4}}+\frac{\lambda}{r}\ln\!\frac{r}{|\lambda|}\right).\label{dd9}
\end{align}

From the expression in Eq.~(\ref{dd9}), it is evident that the geometric parameters, such as the black hole mass $M$, the string cloud parameter $\gamma$, the quantum correction parameter $\hat{\alpha}$, and the perfect fluid parameter $\lambda$, significantly modify the effective potential governing the particle dynamics. As a result, the particle motion is circular orbits, in particular, the innermost stable circular orbits (ISCO) is altered by these.

\begin{figure}[ht!]
\begin{center}
\begin{tabular}{ccc}
\includegraphics[height=5cm]{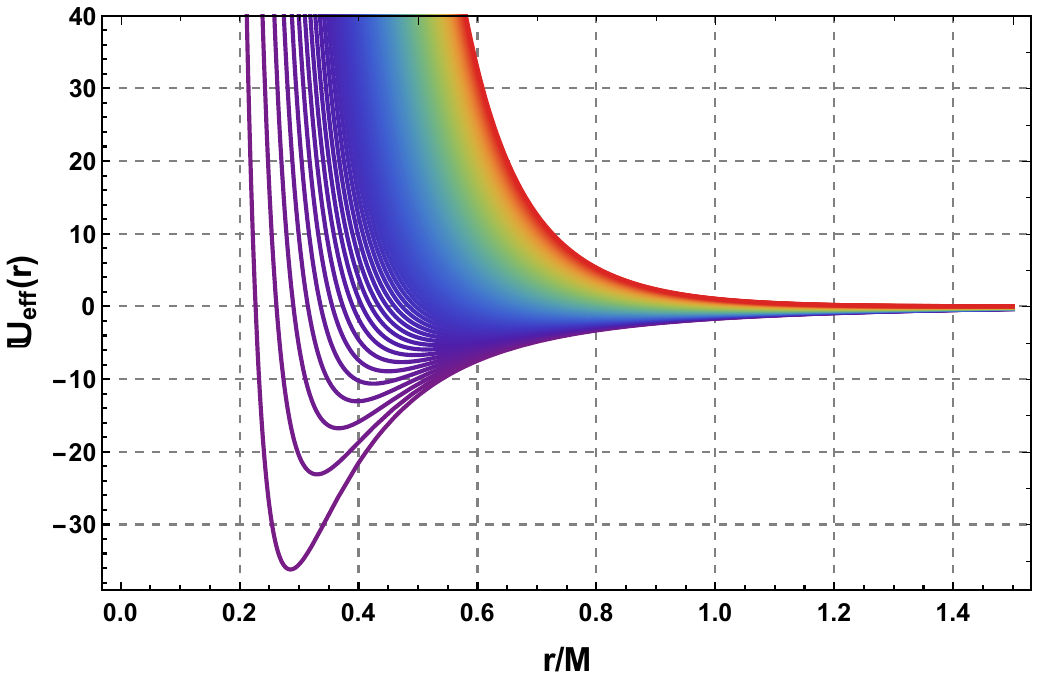} 
\includegraphics[height=5cm]{fig00a.pdf}\\
(a) $\lambda=\gamma=0.1$\\
\includegraphics[height=5cm]{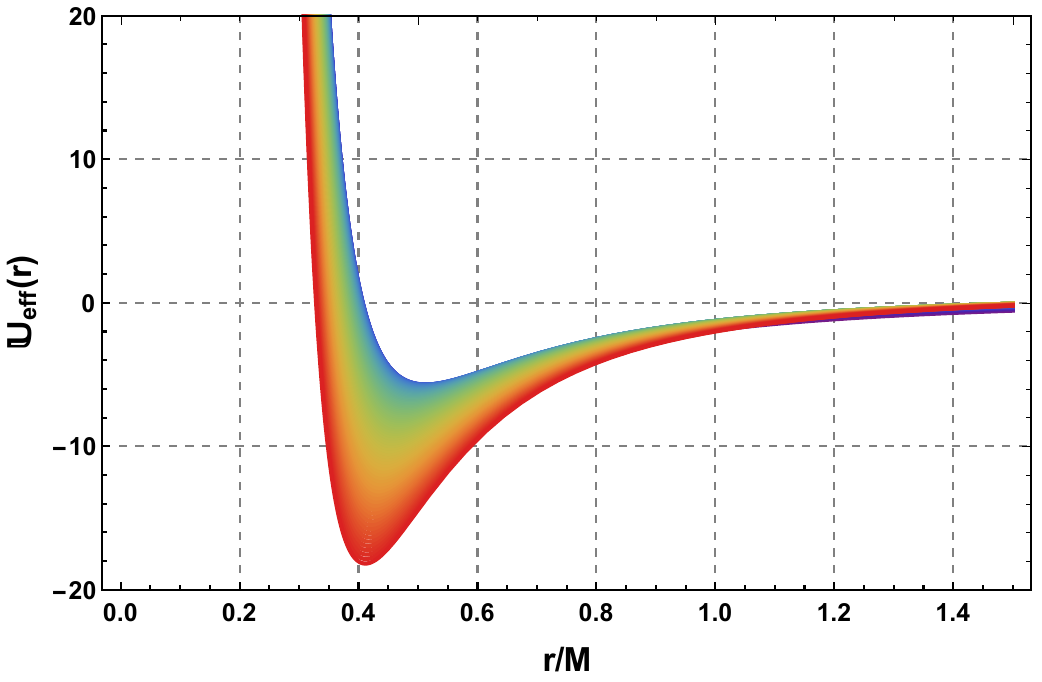} 
\includegraphics[height=5cm]{fig00b.pdf}\\
(b) $\hat{\alpha}=\gamma=0.1$\\
\includegraphics[height=5cm]{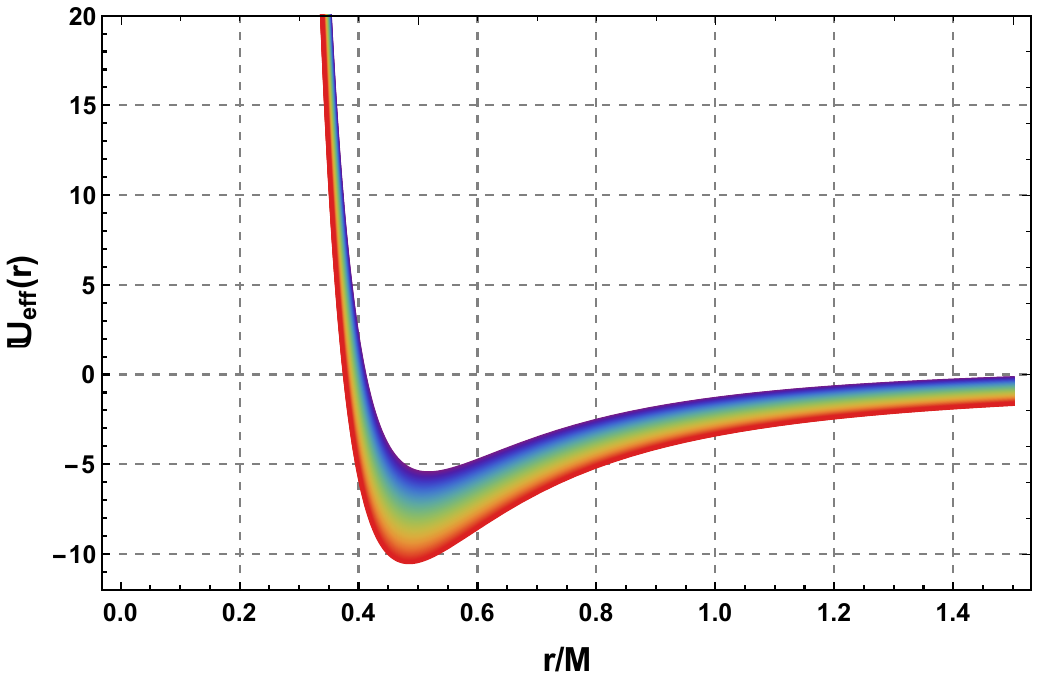} 
\includegraphics[height=5cm]{fig00c.pdf}\\
(c) $\lambda=\hat{\alpha}=0.1$
\end{tabular}
\end{center}
\caption{The behavior of the effective potential determines the particle dynamics as a function of the radial distance. Here, the conserved angular momentum is set to $\mathcal{L}=1$.
\label{fig09}}
\end{figure}

Figure~\ref{fig09} illustrates the radial profile of the effective potential $\mathbb{U}_{\rm eff}(r)$ for massive test particles with fixed angular momentum $\mathcal{L}=1$ and $M=1$, under variations of the deformation parameters $\hat{\alpha}$, $\lambda$, and $\gamma$. In Fig.\ref{fig09}(a), with $\lambda=\gamma=0.1$, increasing the quantum correction parameter $\hat{\alpha}$ significantly modifies the inner structure of the potential, enhancing the height of the centrifugal barrier and shifting the position of the local extrema. This directly affects the location and stability of circular orbits, particularly the ISCO radius. In Fig.\ref{fig09}(b), for $\hat{\alpha}=\gamma=0.1$, variations of the PFDM parameter $\lambda$ lead to moderate deformations of the potential well, slightly altering the depth and radial position of the minimum, which indicates a non-negligible influence of dark matter on bound orbits. Fig.\ref{fig09}(c), with $\lambda=\hat{\alpha}=0.1$, shows that the string cloud parameter $\gamma$ induces a global rescaling of the potential, lowering its asymptotic value and modifying the overall gravitational attraction experienced by test particles. In all cases, the interplay among quantum corrections, string cloud, and PFDM parameters reshapes the potential well structure, thereby changing the conditions for stable and marginally stable circular motion. These modifications have direct implications for the ISCO radius, accretion disk structure, and observational signatures in strong-field regimes.

For circular motions in fixed radius, the conditions $\dot r=0$ and $\ddot r=0$ must be satisfied. These conditions imply the following relation:
\begin{align}
\mathcal{E}^2=\mathbb{U}_{\rm eff} (r).\label{dd10}
\end{align}
And
\begin{equation}
\partial_r U_{\rm eff}=0.\label{dd11}
\end{equation}

Using given potential in (\ref{dd9}), we find the specific angular momentum
\begin{align}
\mathcal{L}^2_{\rm sp}&=\frac{r^3\,f'(r)}{2\,f(r)-r\,f'(r)}\nonumber\\
&=r^2\,\frac{\frac{M}{r} - \frac{2 \hat{\alpha} M^4}{r^4} + \frac{\lambda}{2 r} \left( 1 - \ln\frac{r}{|\lambda|} \right)}{1 - \gamma - \frac{3 M}{r} + \frac{3 \hat{\alpha} M^4}{r^4} + \frac{3 \lambda}{2r} \ln\frac{r}{|\lambda|} - \frac{\lambda}{2r}}.\label{dd12}
\end{align}
And the specific energy as
\begin{align}
\mathcal{E}^2_{\rm sp}&=\frac{f2(r)}{2\,f(r)-r\,f'(r)}\nonumber\\
&=\frac{\left(1 -\gamma- \frac{2M}{r} + \frac{\hat{\alpha} M^{4}}{r^{4}}+\frac{\lambda}{r}\ln\!\frac{r}{|\lambda|}\right)^2}{1 - \gamma - \frac{3 M}{r} + \frac{3 \hat{\alpha} M^4}{r^4} + \frac{3 \lambda}{2r} \ln\frac{r}{|\lambda|} - \frac{\lambda}{2r}}.\label{dd13}
\end{align}

Moreover, for stable circular orbits, the following conditions must be satisfied:
\begin{equation}
\mathcal{E}^2=\mathbb{U}_{\rm eff} (r),\quad \partial_r U_{\rm eff}=0,\quad \partial^2_r U_{\rm eff} \geq 0.\label{dd14}
\end{equation}

For marginally stable circular orbits, the condition $\partial^2_r U_{\rm eff}=0$ results the relation $f(r) f''(r) - 2 \big(f'(r)\big)^2 + \frac{3 f(r) f'(r)}{r} = 0$ which after rearranging results
\begin{widetext}
\begin{align}
&\Bigg( 1 - \gamma - \frac{2M}{r} + \frac{\hat{\alpha} M^4}{r^4} + \frac{\lambda}{r} \ln\frac{r}{|\lambda|} \Bigg)
\Bigg( - \frac{4M}{r} + \frac{20 \hat{\alpha} M^4}{r^4} + \frac{\lambda}{r} \big(2 \ln\frac{r}{|\lambda|} - 3 \big) \Bigg) \nonumber\\
&\quad - 2 \Bigg[ \frac{2M}{r} - \frac{4 \hat{\alpha} M^4}{r^4} + \frac{\lambda}{r} \left( 1 - \ln\frac{r}{|\lambda|} \right) \Bigg]^2
+3 \Bigg( 1 - \gamma - \frac{2M}{r} + \frac{\hat{\alpha} M^4}{r^4} + \frac{\lambda}{r} \ln\frac{r}{|\lambda|} \Bigg)
\Bigg( \frac{2M}{r} - \frac{4 \hat{\alpha} M^4}{r^4} + \frac{\lambda}{r} \left( 1 - \ln\frac{r}{|\lambda|} \right) \Bigg) = 0.\label{dd15}
\end{align}
\end{widetext}

The exact analytical solution of the above polynomial gives us the ISCO radius. Noted that the analytical solution of this polynomial is a quite challenging due to the higher order's in $r$. However, numerical values of ISCO radius can be determined by selecting suitable values of the geometric parameters involved the space-time metric.

\section{Scalar Perturbations of BH}\label{S5}

Scalar field perturbations constitute a fundamental tool for probing the dynamical stability and resonant structure of black hole (BH) spacetimes. By analyzing the propagation of test scalar fields on a fixed gravitational background, one can extract essential information about the geometry through characteristic oscillation spectra, commonly referred to as quasi-normal modes (QNMs). These modes depend exclusively on the parameters defining the BH spacetime and are independent of the initial perturbation, thereby providing a distinctive fingerprint of the underlying geometry \cite{Regge1957,Zerilli1970,Chandrasekhar1984,Berti2009,Konoplya2011}.

In this section, we investigate massless scalar perturbations in the static and spherically symmetric BH background described by the metric function (\ref{fucntion}) where $\gamma$, $\hat{\alpha}$, and $\lambda$ parametrize deviations from the Schwarzschild geometry. Such corrections may arise from modified gravity effects or effective descriptions incorporating quantum or nonlocal contributions. The impact of these terms on the scalar perturbation spectrum is one of the central motivations of the present analysis. For related studies, see, for example, \cite{FA5,FA6,FA7,FA8} and the references therein.

The dynamics of a massless scalar field $\Psi$ minimally coupled to gravity is governed by the covariant Klein--Gordon equation,
\begin{equation}
\frac{1}{\sqrt{-g}}\,\partial_{\mu}\!\left(\sqrt{-g}\,g^{\mu\nu}\,\partial_{\nu}\Psi\right)=0,
\label{KG}
\end{equation}
where $g_{\mu\nu}$ denotes the spacetime metric and $g=\det(g_{\mu\nu})$.

For the line element (\ref{metric})
the metric tensor and its inverse take the diagonal form
\begin{eqnarray}
g_{\mu\nu}&=&\mathrm{diag}\!\left(-f(r),\,f^{-1}(r),\,r^2,\,r^2\sin^2\theta\right),\nonumber\\
g^{\mu\nu}&=&\mathrm{diag}\!\left(-f^{-1}(r),\,f(r),\,r^{-2},\,r^{-2}\sin^{-2}\theta\right),
\end{eqnarray}
with determinant $g=-r^4\sin^2\theta$.

Exploiting the spherical symmetry of the background, we decompose the scalar field as
\begin{equation}
\Psi(t,r,\theta,\phi)=e^{-i\omega t}\,Y_{\ell}^{m}(\theta,\phi)\,\frac{\psi(r)}{r},
\label{ansatz}
\end{equation}
where $\omega$ is the (generally complex) frequency and $Y_{\ell}^{m}$ are the spherical harmonics.

Substituting the ansatz (\ref{ansatz}) into Eq.~(\ref{KG}) and introducing the tortoise coordinate $r_*$ defined by
\begin{equation}
\frac{dr_*}{dr}=\frac{1}{f(r)},
\label{tortoise}
\end{equation}
the radial equation can be cast into a Schr\"odinger-like form,
\begin{equation}
\frac{d^2\psi}{dr_*^2}+\left[\omega^2-V_{\text{s}}(r)\right]\psi=0.
\label{Schrodinger}
\end{equation}

The effective potential governing scalar perturbations is given by
\begin{eqnarray}
V_{\text{s}}(r)&=&f(r)\left[\frac{\ell(\ell+1)}{r^2}+\frac{1}{r}\frac{df(r)}{dr}\right]\nonumber\\
&=&\frac{1}{r^2}\Bigg[1-\gamma-\frac{2M}{r}+\frac{\hat{\alpha} M^{4}}{r^{4}}+\frac{\lambda}{r}\ln\!\frac{r}{|\lambda|}\Bigg]\nonumber\\
&\times&\Bigg\{\ell(\ell+1)+\frac{2M}{r}-\frac{4\hat{\alpha}M^4}{r^4}+\frac{\lambda}{r}\left(1-\ln\!\frac{r}{|\lambda|}\right)\Bigg\}.\quad\quad
\end{eqnarray}

The resulting wave equation (\ref{Schrodinger}) serves as the starting point for computing QNMs using semi-analytical or numerical techniques, such as the WKB approximation or time-domain integration, which will be explored in the subsequent sections.

\begin{figure}[ht!]
\begin{center}
\begin{tabular}{ccc}
\includegraphics[height=5cm]{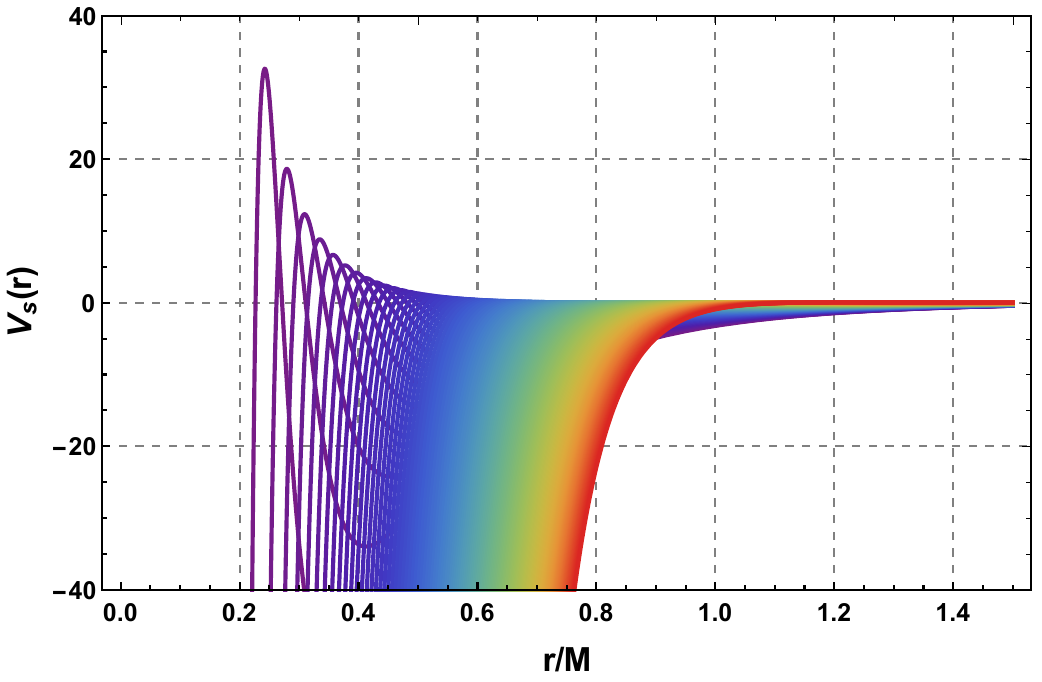} 
\includegraphics[height=5cm]{fig00a.pdf}\\
(a) $\lambda=\gamma=0.1$\\
\includegraphics[height=5cm]{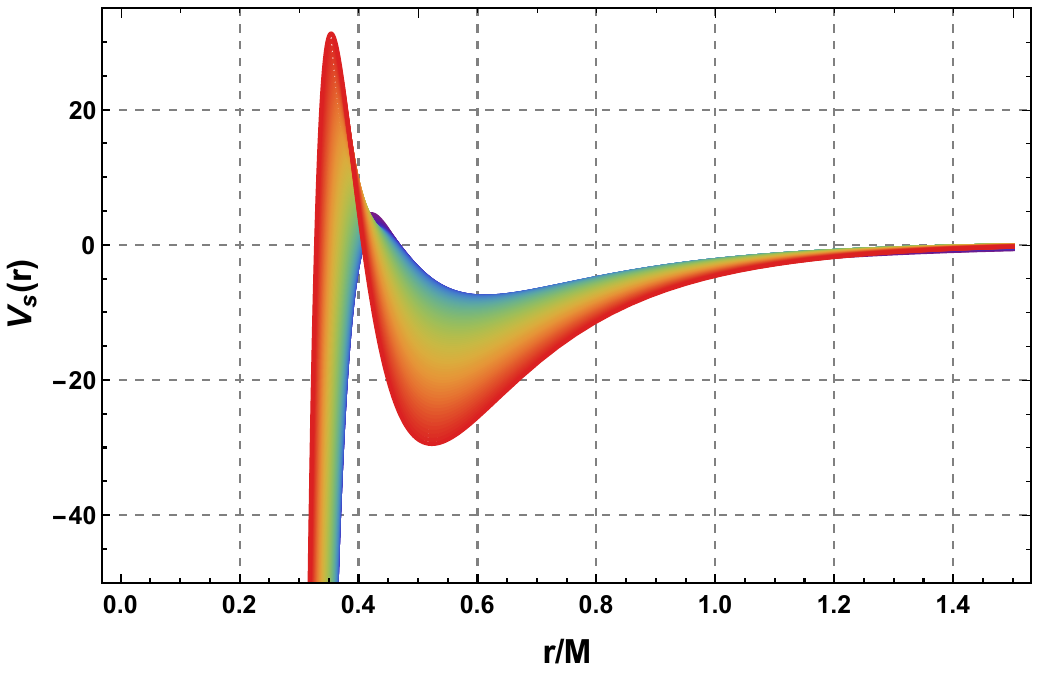} 
\includegraphics[height=5cm]{fig00b.pdf}\\
(b) $\hat{\alpha}=\gamma=0.1$\\
\includegraphics[height=5cm]{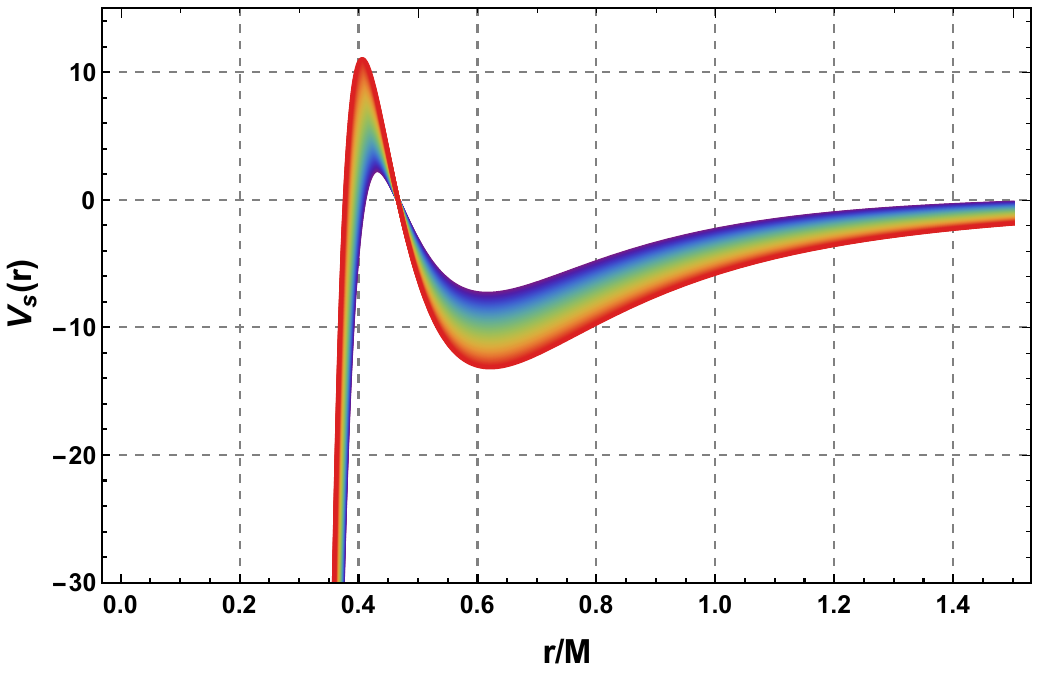} 
\includegraphics[height=5cm]{fig00c.pdf}\\
(c) $\lambda=\hat{\alpha}=0.1$
\end{tabular}
\end{center}
\caption{The behavior of the effective potential governing scalar perturbations as a function of radial distance by varying $\hat{\alpha}, \lambda$ and $\gamma$. Here $\ell=1$.
\label{fig08}}
\end{figure}

Figure~\ref{fig08} displays the radial behavior of the effective potential $V_{\text{s}}(r)$ governing massless scalar perturbations for $\ell=1$ and $M=1$, under variations of the deformation parameters $\hat{\alpha}$, $\lambda$, and $\gamma$ while keeping the remaining parameters fixed. In Fig.\ref{fig08} (a), with $\lambda=\gamma=0.1$, increasing the quantum correction parameter $\hat{\alpha}$ substantially modifies the height and sharpness of the potential barrier in the near-horizon region, leading to a steeper peak and shifting its radial position. This behavior directly influences the quasi-normal mode spectrum, as the barrier characteristics determine the oscillation frequencies and damping rates. In Fig.\ref{fig08} (b), for $\hat{\alpha}=\gamma=0.1$, variations of the PFDM parameter $\lambda$ smooth the potential profile and reduce the barrier height, indicating that the presence of dark matter effectively weakens the confining nature of the potential well. Fig.\ref{fig08} (c), with $\lambda=\hat{\alpha}=0.1$, shows that the string cloud parameter $\gamma$ produces a global suppression of the potential, lowering its amplitude across all radial scales and modifying both the peak structure and asymptotic decay. In all cases, the potential approaches zero at large radial distances, confirming asymptotic flatness and ensuring purely outgoing boundary conditions at spatial infinity. These results demonstrate that quantum corrections primarily affect the strong-field barrier structure, while PFDM and string cloud contributions introduce cumulative large-scale modifications, with direct consequences for the stability analysis and the scalar quasi-normal mode spectrum of the black hole spacetime.

\section{Thermodynamics of BH}\label{S6}

In this section, we study the thermodynamic properties of the black hole solution by deriving key quantities and analyzing their behavior. For related investigations, see, for example, \cite{FA1,FA3,FA4,FA7} and the references therein.

The thermodynamic behavior of the black hole is entirely determined by the properties of the event horizon associated with the metric function $f(r)$ given in Eq.~(\ref{fucntion}). In this section, we derive the black hole mass and Hawking temperature as functions of the horizon radius $r_{+}$, highlighting the effects of the deformation parameters $\gamma$, $\hat{\alpha}$, and $\lambda$.

The event horizon is defined as the largest positive root of the equation
\begin{equation}
f(r_{h})=0.
\label{horizon}
\end{equation}
Using Eq.~(\ref{fucntion}), the horizon condition explicitly reads
\begin{equation}
1-\gamma-\frac{2M}{r_{h}}+\frac{\hat{\alpha} M^{4}}{r_{h}^{4}}
+\frac{\lambda}{r_{h}}\ln\!\left(\frac{r_{h}}{|\lambda|}\right)=0.
\label{horizon_eq}
\end{equation}

The Hawking temperature is determined from the surface gravity $\kappa$ at the event horizon \cite{Bekenstein1973,Hawking1975,Hawking1974},
\begin{eqnarray}
T_{H}&=&\frac{\kappa}{2\pi}
=\frac{f'(r_{h})}{4\pi}\nonumber\\
&=&\frac{1}{4\pi}\left[
\frac{2M}{r_{h}^{2}}
-\frac{4\hat{\alpha}M^{4}}{r_{h}^{5}}
+\frac{\lambda}{r_{h}^{2}}\left(1-\ln\!\frac{r_{h}}{|\lambda|}\right)
\right]
,
\label{Hawking_def}
\end{eqnarray}
where the prime denotes differentiation with respect to $r$. 
This expression clearly demonstrates how the logarithmic correction governed by $\lambda$ and the higher-order mass contribution proportional to $\hat{\alpha}$ modify the standard Hawking temperature. In particular, these terms can significantly affect the thermal behavior of the black hole for small horizon radii, potentially leading to deviations from the classical evaporation scenario. In the Schwarzschild limit, Eq.~(\ref{Hawking_def}) consistently reduces to
\begin{equation}
T_{H}=\frac{1}{4\pi r_{h}},
\end{equation}
thereby confirming the internal consistency of the thermodynamic framework.

For static, spherically symmetric configurations, the entropy satisfies the Bekenstein-Hawking area law \cite{Bekenstein1973,Hawking1975,Hawking1974},
\begin{eqnarray}
S=\frac{A}{4}=\pi r_{h}^{2}.
\end{eqnarray}
The heat capacity at constant pressure can then be written as \cite{Davies1977,Bardeen1973,Wald1994}
\begin{eqnarray}
C_{p}=T_{H}\,\left(\frac{\partial S}{\partial T_{H}}\right).
\end{eqnarray}
Finally, the Gibbs free energy is defined by
\begin{eqnarray}
G=M-T_{H}S.
\end{eqnarray}

\begin{figure}[ht!]
\begin{center}
\begin{tabular}{ccc}
\includegraphics[height=5cm]{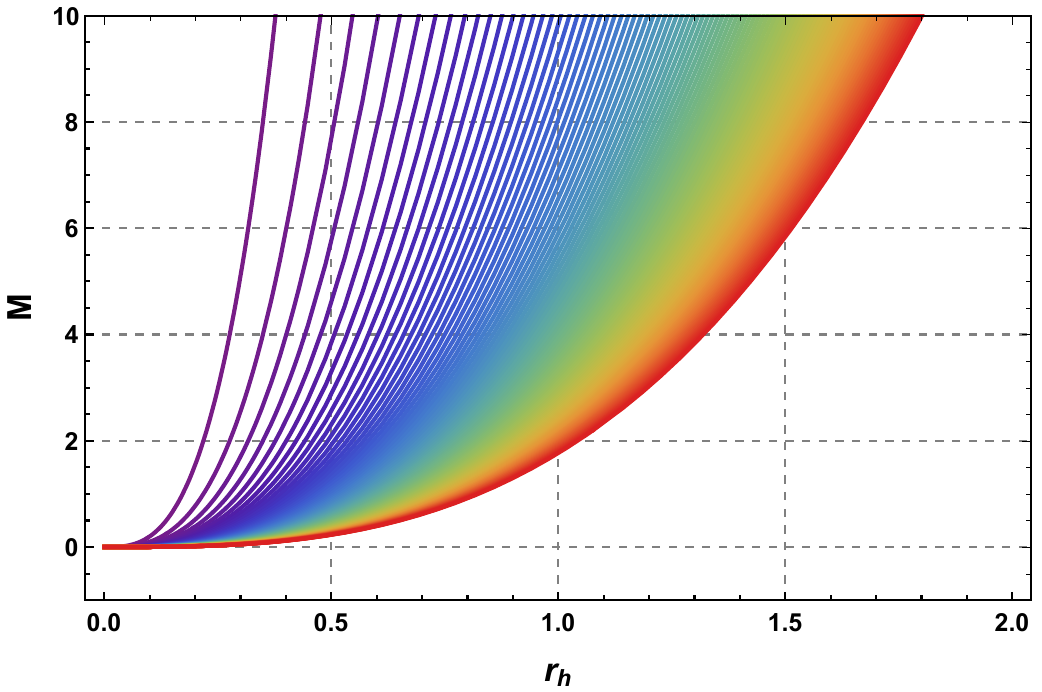} 
\includegraphics[height=5cm]{fig00a.pdf}\\
(a) $\lambda=\gamma=0.1$\\
\includegraphics[height=5cm]{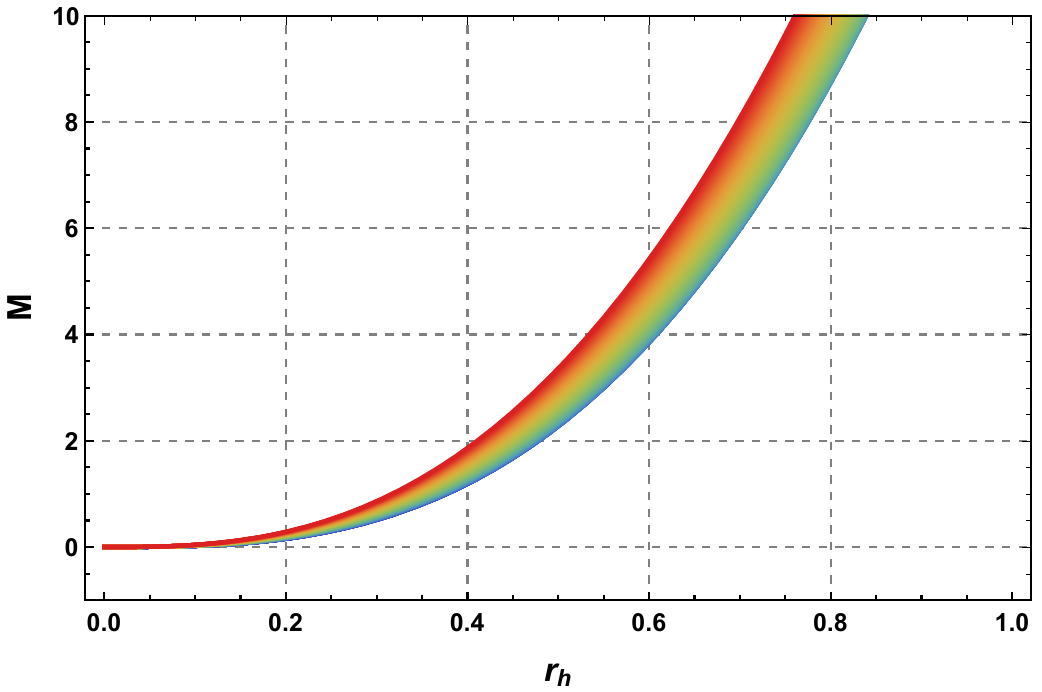} 
\includegraphics[height=5cm]{fig00b.pdf}\\
(b) $\hat{\alpha}=\gamma=0.1$\\
\includegraphics[height=5cm]{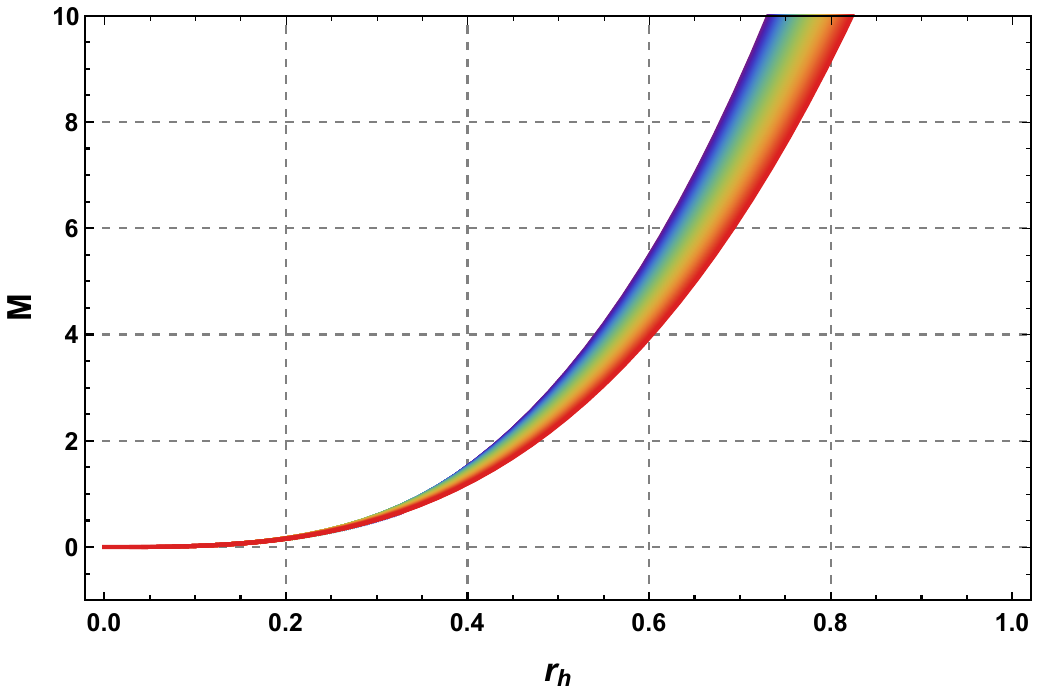} 
\includegraphics[height=5cm]{fig00c.pdf}\\
(c) $\lambda=\hat{\alpha}=0.1$
\end{tabular}
\end{center}
\caption{Behavior of ADS mass as a function of horizon by varying $\hat{\alpha}, \lambda$ and $\gamma$.
\label{fig04}}
\end{figure}

Figure~\ref{fig04} illustrates the dependence of the black hole mass $M$ on the event horizon radius $r_h$ for different values of the deformation parameters $\hat{\alpha}$, $\lambda$, and $\gamma$. In Fig.\ref{fig04}(a), where $\lambda=\gamma=0.1$, variations of the quantum correction parameter $\hat{\alpha}$ significantly modify the mass--horizon relation, especially in the small-$r_h$ regime, indicating that quantum effects become dominant near the horizon scale and allow black hole solutions with smaller masses for a given horizon radius. Fig.\ref{fig04}(b), corresponding to $\alpha=\gamma=0.1$, shows that the perfect fluid dark matter parameter $\lambda$ introduces a comparatively weaker correction, slightly shifting the $M$--$r_h$ curve while preserving its overall monotonic behavior. In contrast, Fig.\ref{fig04}(c) demonstrates that the string cloud parameter $\gamma$, for $\lambda=\alpha=0.1$, induces a more pronounced global deformation of the mass function, reflecting its direct contribution to the effective gravitational field and modifying the black hole mass at all horizon scales. These results indicate that quantum corrections primarily affect the near-horizon mass structure, whereas the string cloud and PFDM parameters govern large-scale deviations from the classical Schwarzschild relation, with important consequences for black hole thermodynamics and stability.

\begin{figure}[ht!]
\begin{center}
\begin{tabular}{ccc}
\includegraphics[height=5cm]{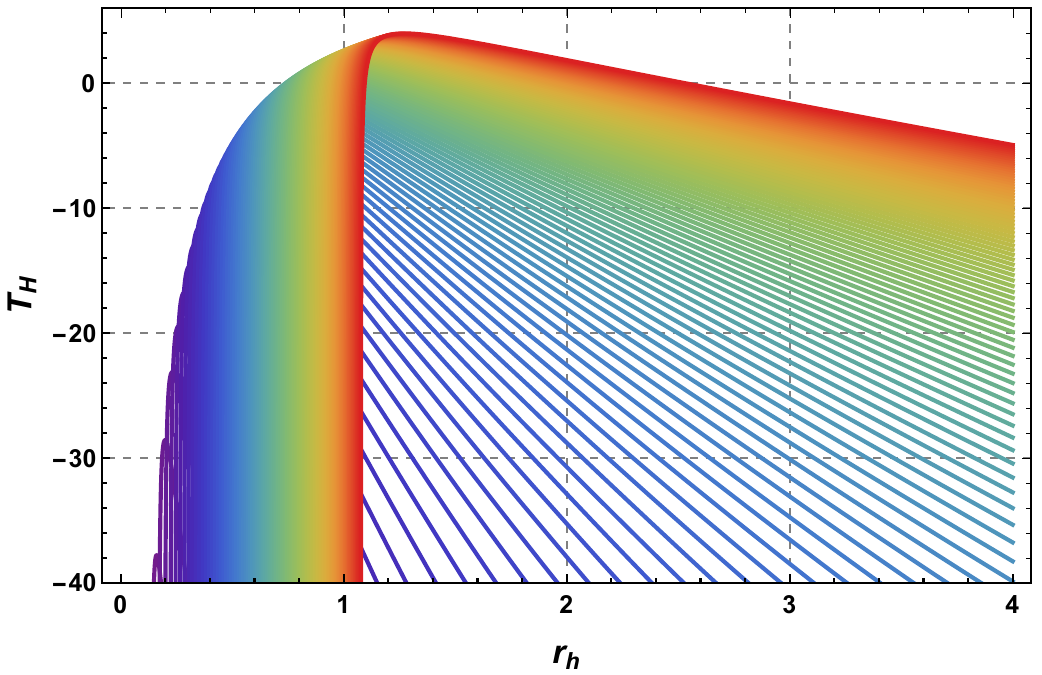} 
\includegraphics[height=5cm]{fig00a.pdf}\\
(a) $\lambda=\gamma=0.1$\\
\includegraphics[height=5cm]{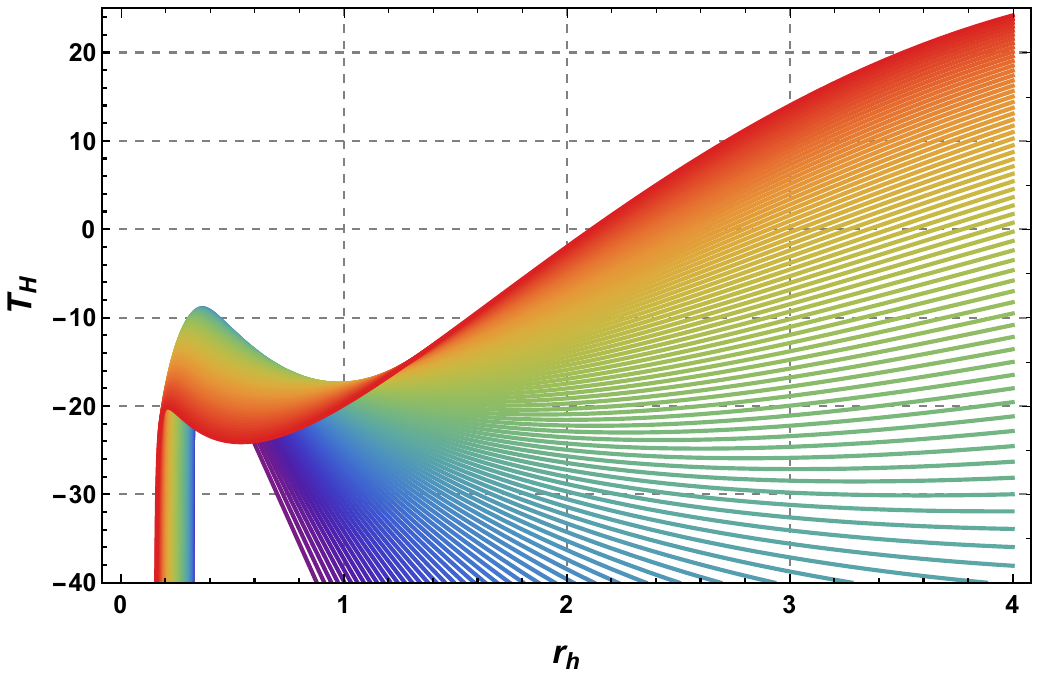} 
\includegraphics[height=5cm]{fig00b.pdf}\\
(b) $\hat{\alpha}=\gamma=0.1$\\
\includegraphics[height=5cm]{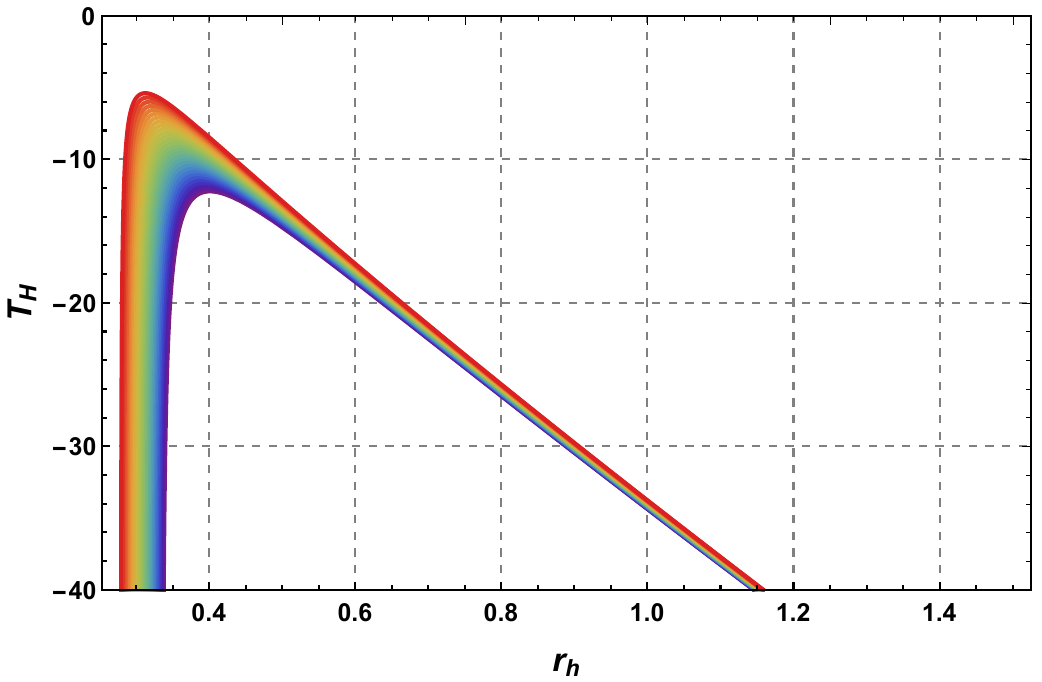} 
\includegraphics[height=5cm]{fig00c.pdf}\\
(c) $\lambda=\hat{\alpha}=0.1$
\end{tabular}
\end{center}
\caption{Behavior of the Hawking Temperature as a function of horizon by varying $\hat{\alpha}, \lambda$ and $\gamma$.
\label{fig05}}
\end{figure}

Figure~\ref{fig05} illustrates the behavior of the Hawking temperature $T_H$ as a function of the horizon radius $r_h$ for different choices of the model parameters $\alpha$, $\lambda$, and $\gamma$. In Fig.\ref{fig05}(a), with $\lambda=\gamma$ fixed, varying $\alpha$ mainly shifts the overall magnitude of $T_H$ while preserving its qualitative profile, indicating that this parameter controls the strength of quantum or higher-order corrections without altering the underlying thermodynamic structure. Fig.\ref{fig05}(b) shows that changes in $\lambda$, with $\alpha=\gamma$ held fixed, significantly modify the slope of $T_H(r_h)$ at larger radii and may induce local extrema, suggesting a nontrivial impact on the thermal response of the black hole and the possible emergence of thermodynamically distinct regimes. In contrast, Fig.\ref{fig05}(c) reveals that variations of $\gamma$, for fixed $\lambda=\alpha$, primarily affect the small-horizon region, leading to a pronounced maximum of $T_H$ followed by a monotonic decrease, which is typically associated with the existence of a remnant-like phase or a stabilization mechanism at short scales. 

\begin{figure}[ht!]
\begin{center}
\begin{tabular}{ccc}
\includegraphics[height=5cm]{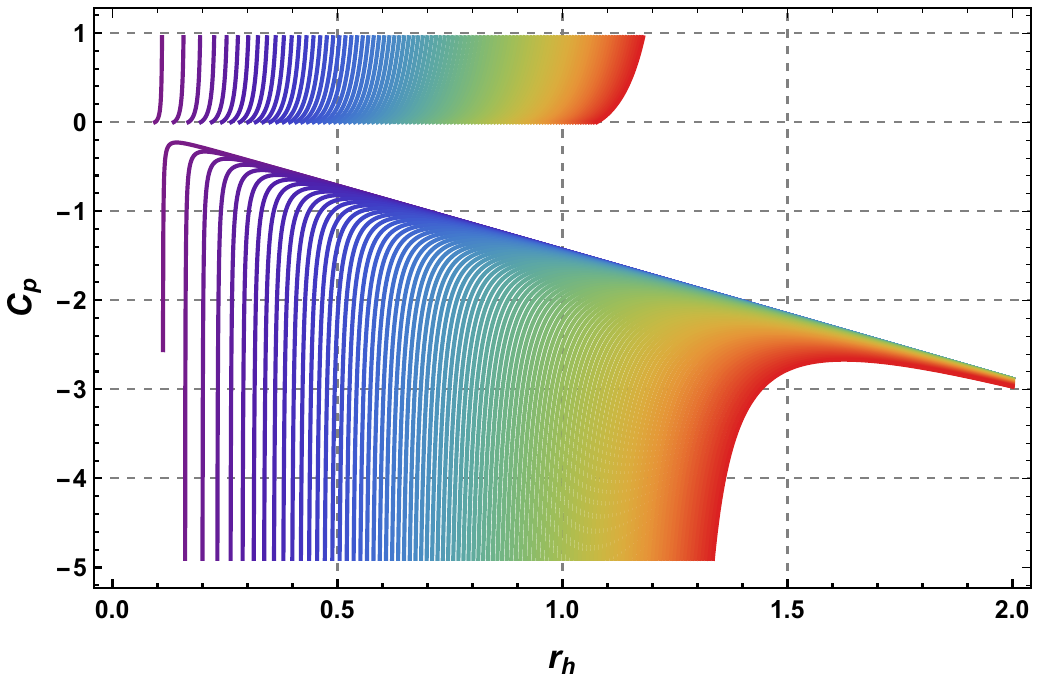} 
\includegraphics[height=5cm]{fig00a.pdf}\\
(a) $\lambda=\gamma=0.1$\\
\includegraphics[height=5cm]{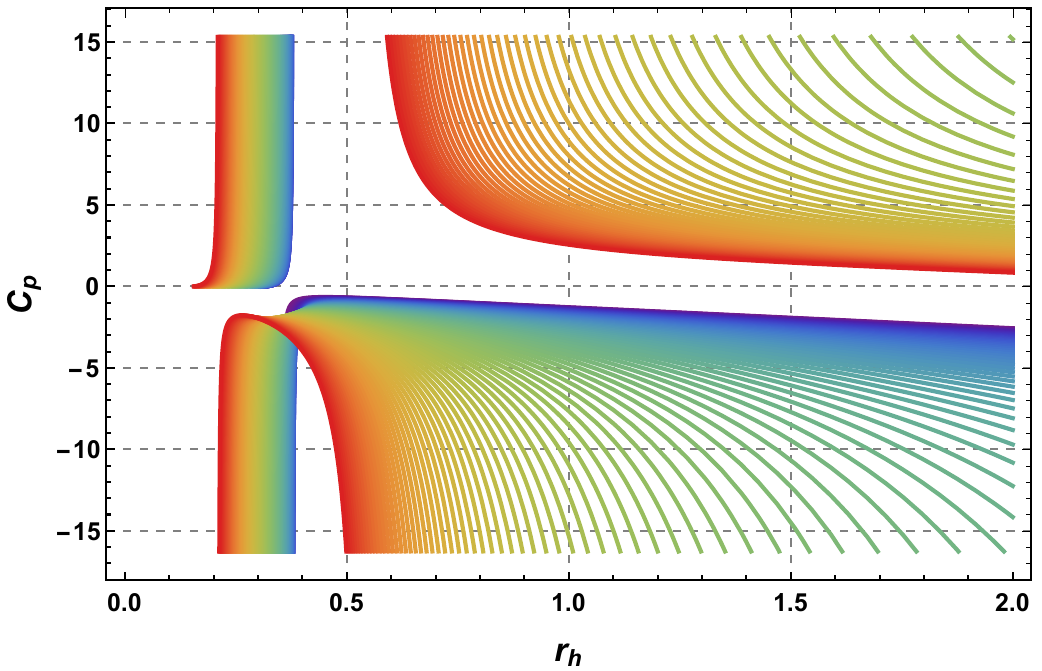} 
\includegraphics[height=5cm]{fig00b.pdf}\\
(b) $\hat{\alpha}=\gamma=0.1$\\
\includegraphics[height=5cm]{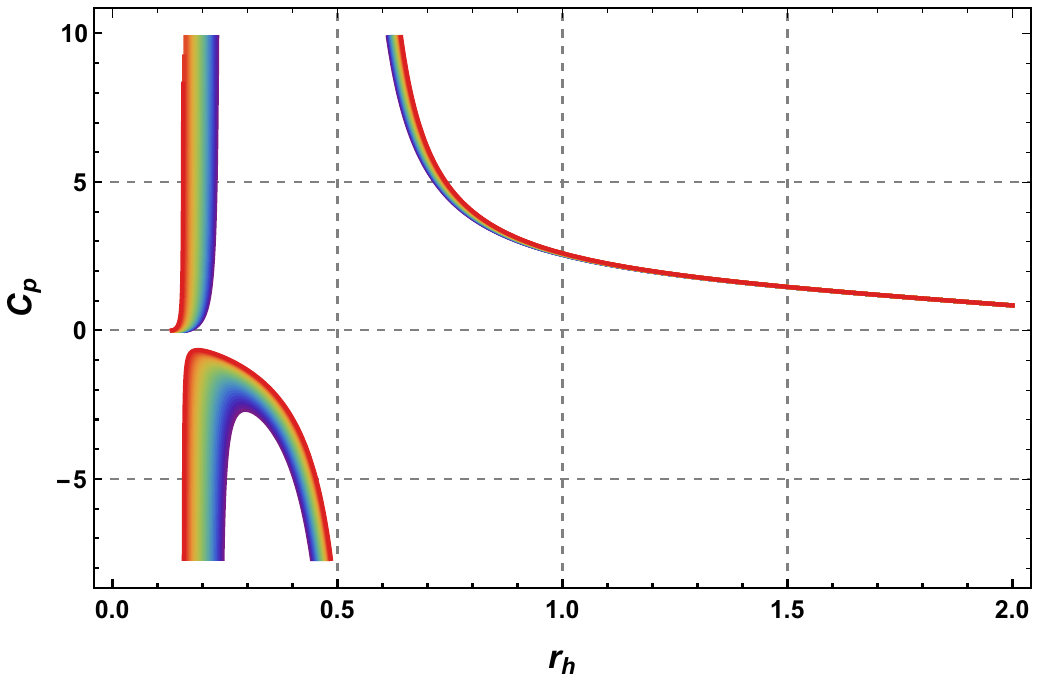} 
\includegraphics[height=5cm]{fig00c.pdf}\\
(c) $\lambda=\hat{\alpha}=0.1$
\end{tabular}
\end{center}
\caption{Illustration of specific heat capacity as a function of horizon by varying $\hat{\alpha}, \lambda$ and $\gamma$. 
\label{fig06}}
\end{figure}

Figure~\ref{fig06} displays the heat capacity $C_Q$ as a function of the horizon radius $r_h$ for different values of the parameters $\alpha$, $\lambda$, and $\gamma$. In Fig.\ref{fig06}(a), where $\lambda=\gamma$ are fixed, variations in $\alpha$ modify both the magnitude and the location of divergences in $C_Q$, indicating the presence of phase transition points associated with changes in thermodynamic stability. Fig.\ref{fig06}(b) shows that the parameter $\lambda$, with $\alpha=\gamma$ held constant, plays a crucial role in shaping the asymptotic behavior of the heat capacity, leading to extended regions with positive $C_Q$ that signal locally stable black hole configurations, as well as negative branches corresponding to unstable phases. In Fig.\ref{fig06}(c), varying $\gamma$ while fixing $\lambda=\alpha$ mainly affects the small-horizon regime, where sharp transitions between positive and negative heat capacity emerge, suggesting a sensitive dependence of short-scale thermodynamics on this coupling.

\begin{figure}[ht!]
\begin{center}
\begin{tabular}{ccc}
\includegraphics[height=5cm]{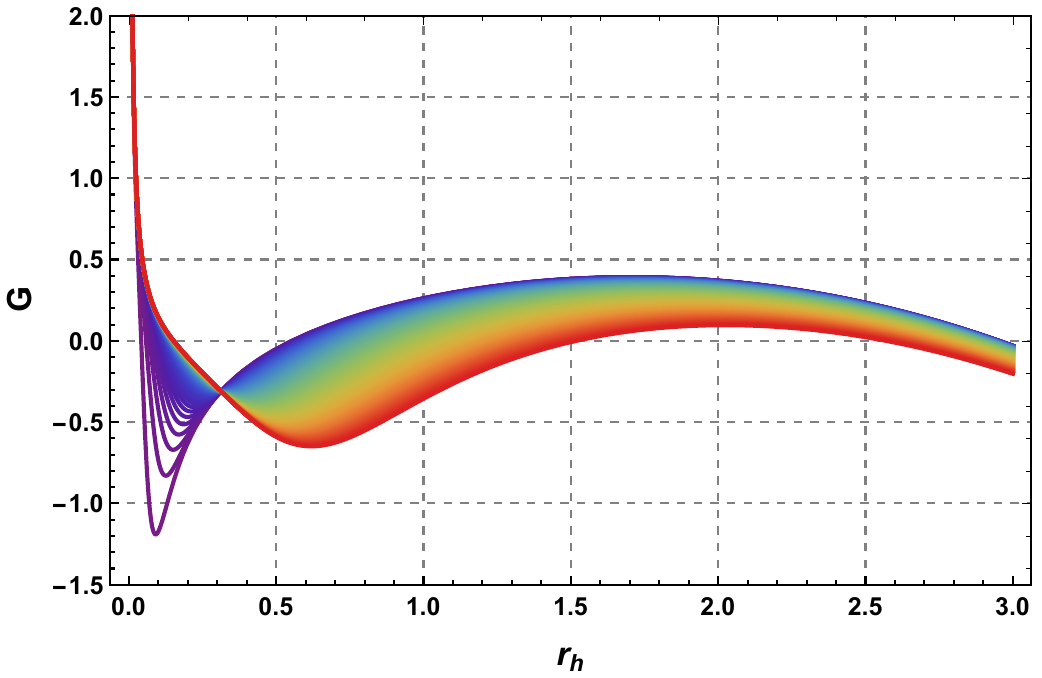} 
\includegraphics[height=5cm]{fig00a.pdf}\\
(a) $\lambda=\gamma=0.1$\\
\includegraphics[height=5cm]{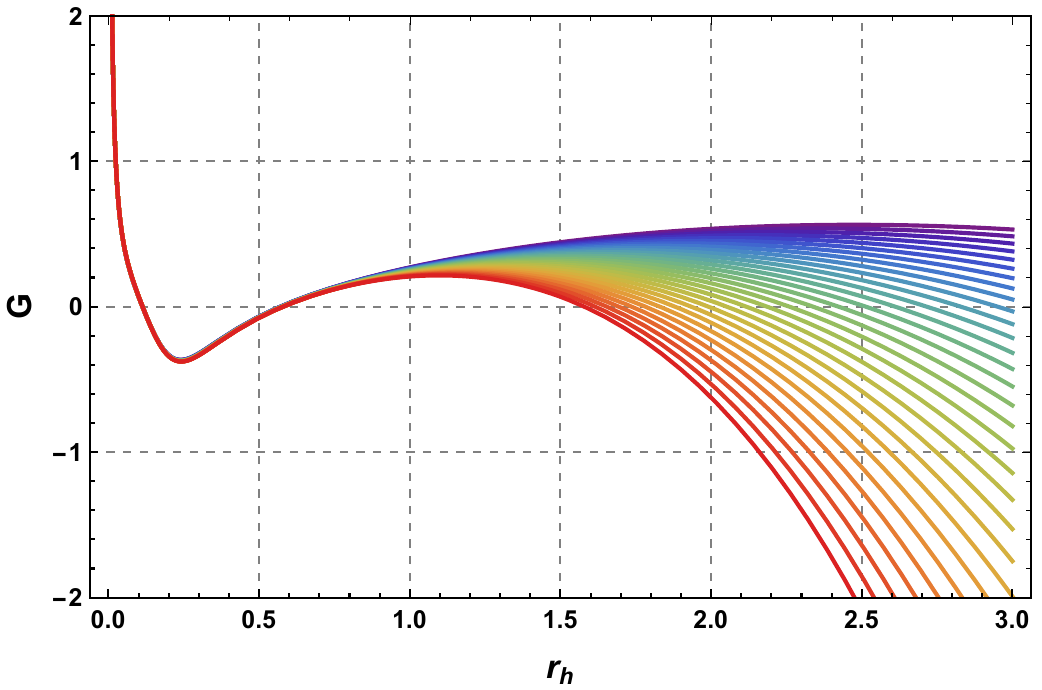} 
\includegraphics[height=5cm]{fig00b.pdf}\\
(b) $\hat{\alpha}=\gamma=0.1$\\
\includegraphics[height=5cm]{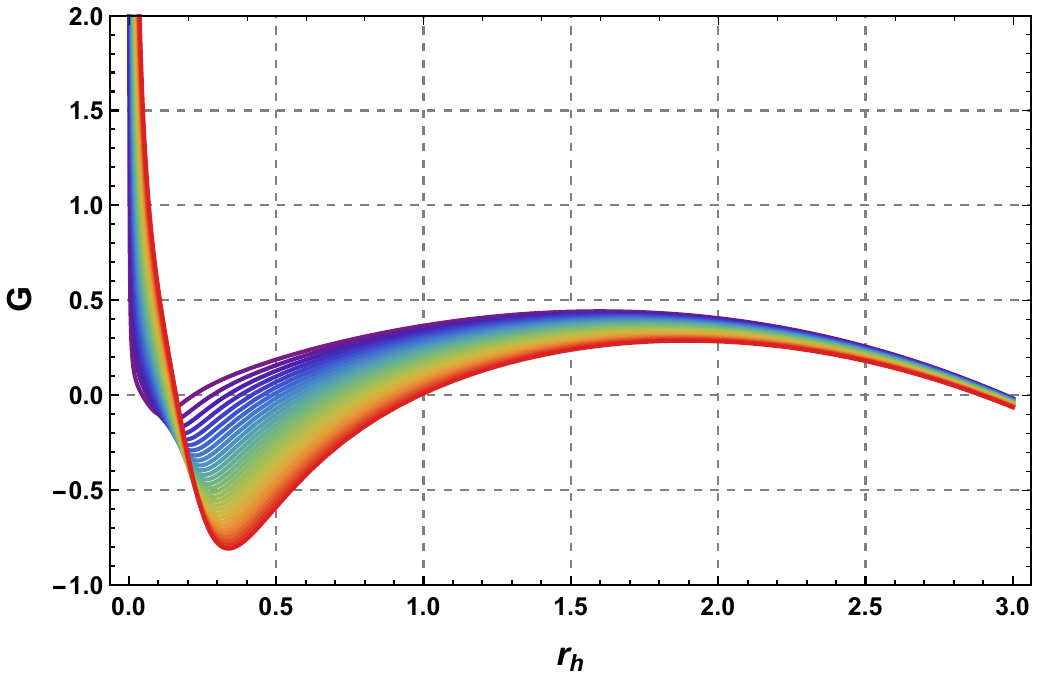} 
\includegraphics[height=5cm]{fig00c.pdf}\\
(c) $\lambda=\hat{\alpha}=0.1$
\end{tabular}
\end{center}
\caption{Illustration of Gibb's free energy as a function of horizon by varying $\hat{\alpha}, \lambda$ and $\gamma$. 
\label{fig07}}\label{fig08b}
\end{figure}

Figure~\ref{fig07} shows the behavior of the Gibbs free energy $G$ as a function of the horizon radius $r_h$ for different values of the parameters $\alpha$, $\lambda$, and $\gamma$. In Fig.\ref{fig07}(a), where $\lambda=\gamma$ are fixed, increasing $\alpha$ modifies the depth and position of the minimum of $G$, indicating that the system admits thermodynamically preferred black hole states characterized by negative free energy over a finite range of horizon radii. 
Fig.\ref{fig07}(b) reveals that variations of $\lambda$, with $\alpha=\gamma$ held constant, strongly affect the large-$r_h$ regime, producing a transition from negative to positive Gibbs free energy and signaling a change in the globally favored phase. In Fig.\ref{fig07}(c), the parameter $\gamma$ mainly influences the small-horizon behavior, shifting the location of the free-energy minimum and altering the onset of thermodynamic preference. 

\section{Concluding Remarks}\label{S7}

In this study, we concisely summarize and analyze the physical content of the background configuration and the associated geometrical response within a single coherent framework. The resulting spacetime represents a quantum-corrected Oppenheimer-Snyder BH embedded simultaneously in a cloud of strings and a PFDM background, thus incorporating three independent and physically well-motivated deformations with respect to the Schwarzschild geometry. Also, the quantum correction contribution, governed by the dimensionless parameter $(\hat{\alpha})$, generates higher-order inverse-radial terms that dominate in the strong-field domain, leading to pronounced modifications of the near-horizon geometry while leaving the asymptotic behavior effectively unchanged. %in the limit $(\hat{\alpha}\to 0)$, the metric continuously reduces to its classical form, consistently reproducing the Schwarzschild or Letelier-type solutions. The cloud of strings parameter $(\gamma)$ induces a global geometrical deformation by shifting the asymptotic value of the lapse function and introducing a solid-angle deficit, thereby modifying the gravitational potential over all radial scales, in agreement with established Letelier BH constructions. By contrast, the PFDM parameter $(\lambda)$ produces a logarithmic correction that predominantly influences intermediate and large radial regions, capturing the effect of an extended dark matter distribution rather than short-distance quantum corrections. Its contribution vanishes smoothly for $(\lambda=0)$, ensuring consistency with vacuum string-cloud spacetimes. Also, the joint influence of these parameters is explicitly encoded in the lapse function and curvature invariants: $(\hat{\alpha})$ determines the magnitude and structure of curvature in the central region, $(\lambda)$ regulates the curvature profile away from the core, and $(\gamma)$ rescales the global curvature strength. The Ricci and Kretschmann scalars indicate that the spacetime remains asymptotically flat and physically regular outside the horizon, while their divergence as $(r\to 0)$ confirms the persistence of a central curvature singularity whose intensity depends on the combined effects of quantum corrections, string clouds, and PFDM. In this case, the regularity of these invariants at the event horizon and the existence of a well-defined solution of $(f(r)=0)$ ensure that the singularity is always enclosed by an event horizon within the admissible parameter ranges, thereby maintaining the BH nature of the solution and supporting cosmic censorship. Relative to previously investigated Schwarzschild, Letelier, or PFDM BHs, the present construction offers a more comprehensive and physically consistent framework in which quantum gravity effects govern the strong-field regime, while string clouds and dark matter control large-scale deviations, with direct consequences for horizon properties, geodesic dynamics, and observable phenomena such as shadows and gravitational lensing. 

The preceding investigation of null geodesics and optical characteristics demonstrates that the simultaneous inclusion of quantum corrections, a string cloud, and PFDM produces nontrivial and distinguishable changes in the astrophysical observables of the BH relative to the canonical Schwarzschild geometry. The effective potential explicitly indicates that unstable photon trajectories, which determine the photon sphere and delineate the shadow edge, exhibit strong sensitivity to the spacetime deformation parameters. Departures from classical regime emphasize distinct physical contribution: quantum-correction parameter $\hat{\alpha}$ mainly influences the near-horizon geometry by altering the strong-field curvature, producing a measurable displacement of the effective potential peak and hence a modification of the photon sphere radius. This result indicates that quantum contributions, although negligible at asymptotic distances, may generate observable effects at horizon scales, particularly in the shadow size. Conversely, the string cloud parameter $\gamma$ induces a global deficit in the geometry, uniformly decreasing the effective potential across all radii and diminishing the asymptotic metric function, which directly rescales the shadow radius by the factor $(1-\gamma)^{1/2}$. The PFDM parameter $\lambda$, through its logarithmic term, introduces a weaker but long-range modification that becomes significant at intermediate and large distances, softening the potential barrier and slightly shifting the photon sphere position and critical impact parameter. Relative to the Schwarzschild case and other models without dark matter, the present analysis predicts either an increase or a decrease in the photon sphere and shadow radius depending on the combined influence of $\hat{\alpha}$, $\gamma$, and $\lambda$. In this case, these deviations are especially relevant for high-resolution measurements by the Event Horizon Telescope, where modest variations in the shadow radius or photon ring morphology can act as diagnostics of deviations from GR and of ambient matter effects. In this context, the comparison with established limits confirms that the Schwarzschild BH solution is recovered when additional fields are absent, while the incorporation of quantum corrections, string clouds, and PFDM yields potentially detectable signatures in strong gravitational lensing and BH shadows, providing a viable pathway for confronting such effects with present and forthcoming observational data. 

The investigation of geodesic motion and scalar field perturbations demonstrates how the deviation parameters $(\gamma)$, $(\hat{\alpha})$, and $(\lambda)$ generate observable imprints on both classical trajectories and wave dynamics in the vicinity of the BH. From the viewpoint of test particle motion, the structure of the effective potential determines the existence and stability properties of circular orbits, thereby controlling the position of the ISCO and, as a result, the inner boundary of accretion disks. Relative to the Schwarzschild spacetime, the inclusion of the constant shift $(\gamma)$ leads to a rescaling of the asymptotic gravitational potential, altering the depth of the potential well and displacing the radii associated with stable orbital motion. In this case, the higher-order contribution proportional to $(\hat{\alpha} M^{4}/r^{4})$ becomes significant in the strong-field regime, introducing corrections that are not present in GR and that may either enhance or suppress the stability of circular orbits near the event horizon, depending on its sign and magnitude. In parallel, the logarithmic term governed by $(\lambda)$, which can be linked to nonlocal or dark-sector effects, induces a slowly varying modification that affects particle dynamics across an extended radial domain, producing discernible deviations in the asymptotic behavior of the effective potential. Also, in the limiting case $(\gamma \to 0)$, $(\hat{\alpha} \to 0)$, and $(\lambda \to 0)$, the effective potential continuously approaches the Schwarzschild form, restoring the standard ISCO radius and validating the internal coherence of the theoretical framework. The results observed in the geodesic structure are consistently illustrated in the properties of scalar perturbations since the scalar effective potential is derived from the same underlying metric function. Also, departures from the Schwarzschild BH geometry modify both the height and profile of the potential barrier that governs wave propagation, thereby influencing the quasi-normal mode spectrum through shifts in oscillation frequencies and decay rates. In particular, variations in the maximum of the scalar potential are expected to be associated with changes in the photon sphere and unstable null geodesics, supporting the established correspondence between eikonal QNMs and circular photon orbits. Consequently, while classical observables such as accretion disk characteristics and shadow radius probe the spacetime geometry through geodesic motion, scalar perturbations offer an independent dynamical probe via ringdown signatures. Also, the overall consistency of these findings, together with the recovery of the Schwarzschild limit, demonstrates the robustness of the model and indicates that the parameters $(\gamma)$, $(\hat{\alpha})$, and $(\lambda)$ represent genuine strong- and intermediate-field corrections that may be constrained using present and forthcoming astrophysical data.

The thermodynamic investigation indicates that the combined influence of the quantum deformation parameter $\hat{\alpha}$, the PFDM parameter $\lambda$, and the string cloud parameter $\gamma$ produces pronounced departures from the classical Schwarzschild BH behavior, especially within the small-horizon sector. The mass-horizon relation shows that $\hat{\alpha}$ mainly controls near-horizon corrections, permitting smaller BH masses for a fixed $r_h$, which agrees with expectations from higher-order quantum effects that weaken the gravitational interaction at short distances. Also, the PFDM contribution associated with $\lambda$ generates logarithmic corrections that build up slowly as $r_h$ increases, leading to moderate yet sustained deviations from the Schwarzschild mass curve while maintaining monotonic behavior. In this case, the string cloud parameter $\gamma$ induces a global modification of the spacetime geometry, testing the effective gravitational potential across all length scales and consequently displacing the mass function throughout the full horizon domain. These differentiated impacts are also evident in the Hawking temperature, where $\hat{\alpha}$ changes the overall magnitude of $T_H$ without affecting its qualitative profile, whereas $\lambda$ strongly influences the large-radius thermal behavior and may generate extrema linked to competing thermodynamic phases. Also, the parameter $\gamma$ mainly governs the short-distance regime, producing a maximum in $T_H$ followed by a decreasing branch, which suggests the possible formation of a remnant-like state and a deviation from the standard evaporation picture. This behavior is supported by the heat capacity, whose divergences identify second-order phase transitions separating stable from unstable configurations; in this context, $\hat{\alpha}$ fixes the positions of the critical points, $\lambda$ enlarges the region of positive heat capacity at large $r_h$, and $\gamma$ enhances stability changes in the small-horizon region. In this case, the Gibbs free energy further substantiates this combined behavior, demonstrating that negative free-energy minima arise over finite horizon ranges, signaling thermodynamically favored BH states whose position and depth depend sensitively on the deformation parameters. Crucially, in the simultaneous limit $\hat{\alpha}\to0$, $\lambda\to0$, and $\gamma\to0$, all thermodynamic quantities continuously recover their Schwarzschild limits, confirming the internal consistency of the framework. In this context, these results show that quantum effects dominate short-scale thermodynamics, whereas string cloud and PFDM contributions regulate large-scale stability and phase structure, yielding a thermodynamic landscape substantially richer than that predicted by classical GR.

\footnotesize

\section*{Acknowledgments}
F.A. acknowledges the Inter University Centre for Astronomy and Astrophysics (IUCAA), Pune, India for granting visiting associateship.

\section*{Data Availability Statement}
There are no new data associated with this article.

\end{document}